\documentclass[hidelinks,11pt]{article}
\usepackage{natbib,amsmath,amsthm,amssymb,color,graphicx}
\usepackage{times}
\usepackage{epstopdf}
\usepackage{multirow}
\usepackage{hyperref}

\usepackage[usenames,dvipsnames]{xcolor}

\newcommand{\Det}[1]{|#1|}

\newcommand{\Betadis}[1]{\mathcal{B}\left(#1\right)}
\newcommand{\Betafun}[1]{B (#1)}
 \newcommand{\indic}[1]{\mathbb{I}{\{#1\}}}

\newcommand{\nfactrue}{\rtrue}
\newcommand{\Uniform}[1]{\mathcal{U}\left[#1\right]}
\newcommand{\Gammainv}[1]{\mathcal{G}^{-1} (#1)}

\newcommand{\Real}{\mathbb{R}}
\newcommand{\Prob}[1]{{\rm P} (#1)}    \newcommand{\piL}{\pi}
\newcommand{\piLv}{{\boldsymbol{\piL}}} \newcommand{\stick}{\nu}
\newcommand{\betaTau}{\beta}
\newcommand{\Student}[2]{t _{#1} (#2 )}

\newcommand{\diracarg}[1]{\delta_{\{#1\}}}
\newcommand{\Ew}[1]{\mathbb{E}(#1)}
\newcommand{\tauv}{{\mathbf{\boldsymbol{\tau}}}}
\newcommand{\Gammad}[1]{ \mathcal{G}(#1)}

\newcommand{\Fd}[1]{\mbox{\rm F} (#1)}
\newcommand{\Var}{\mathbb{V}}
\newcommand{\V}[1]{\Var (#1 )}
\newcommand{\ym}{{\mathbf y}}
\newcommand{\betav}{\boldsymbol{\beta}} \newcommand{\Vary}{\boldsymbol{\Omega}}         \newcommand{\Vare}{\boldsymbol{\Sigma}}      \newcommand{\facm}{{\mathbf f}}
\newcommand{\error}{\epsilon}             \newcommand{\errorm}{\boldsymbol{\error}} \newcommand{\Normult}[2]{ \mathcal{N} _{#1}\left(#2\right)}
\newcommand{\bfz}{{\mathbf{0}}}
\newcommand{\Diag}[1]{\mbox{\rm Diag}\!\left(#1\right)} \newcommand{\identm}{{\mathbf I}}       \newcommand{\Normal}[1]{ \mathcal{N} \left(#1\right)}

\newcommand{\taucol}{\theta}
\newcommand{\aalpha}{a^\alpha}
\newcommand{\balpha}{b^\alpha}

\newcommand{\acol}{a^\taucol}
\newcommand{\bcol}{b^\taucol}
\newcommand{\ccol}{c^\taucol}
\newcommand{\bkappa}{b^\kappa}
\newcommand{\ckappa}{c^\kappa}
\newcommand{\bsigma}{b^\sigma}
\newcommand{\csigma}{c^\sigma}
\newcommand{\rtrue}{{H_0}}
\newcommand{\maxpar}[1]{#1_{\mbox{{\rm \tiny max}}}}

\newcommand{\cdefl}{c^\nu}
\newcommand{\Edefl}{E^\nu}
\newcommand{\load}{\beta}                 \newcommand{\facload}{\boldsymbol{\load}}                       
\newcommand{\facmt}[1]{\facm_{#1}}
\newcommand{\dimy}{m}                     \newcommand{\dimmat}[2]{#1\times #2}  \newcommand{\tauglob}{\kappa}
\newcommand{\idiov}{\sigma^2}             

\newcommand{\defl}{\nu_0}

   \newcommand{\citeSec}[2]{Section~\ref{#2}}

\newtheorem{proposition}{\bf Proposition}[section]

\newtheorem{definition}{\bf Definition}[section]
\newtheorem{alg}{\bf Algorithm}

\begin{document}
	
\title{Generalized Cumulative Shrinkage Process Priors with Applications to
 Sparse Bayesian Factor Analysis}

\author{Sylvia Fr\"uhwirth-Schnatter\footnote{Department of Finance, Accounting, and Statistics, WU Vienna University of Economics and Business, Austria. Email: {\tt sfruehwi@wu.ac.at}. This paper is one contribution out of 15 to a theme issue \lq \lq Bayesian inference: challenges, perspectives, and prospects\rq\rq\ forthcoming in {\em Philosophical Transactions of the Royal Society, A}, DOI: 10.1098/rsta.2022.0148.}}

\maketitle

\begin{abstract}
 The paper  discusses shrinkage priors which
impose increasing shrinkage in a sequence of parameters.
We review
the cumulative shrinkage process (CUSP) prior of \cite{leg-etal:bay}, which is a
spike-and-slab shrinkage prior where the spike probability is stochastically increasing
 and constructed from the stick-breaking
representation of a Dirichlet process prior. As a first contribution, this CUSP
prior is extended
by involving arbitrary stick-breaking representations arising from
beta distributions.
As a second contribution, we prove that exchangeable spike-and-slab priors,
which are popular
and widely used in sparse Bayesian factor analysis, can be represented as
a finite generalized  CUSP prior, which is easily
obtained from the decreasing order statistics
of the slab probabilities. Hence, exchangeable spike-and-slab shrinkage priors
imply increasing shrinkage as the column index in the loading matrix increases, without imposing
explicit order constraints on the slab probabilities.
An application to  sparse Bayesian factor analysis
illustrates the usefulness of the findings of this paper.
  A new exchangeable  spike-and-slab shrinkage prior based on the triple gamma prior of
\cite{cad-etal:tri} is introduced and shown
 to be  helpful
for estimating the unknown number of factors in a simulation study.

\end{abstract} 
\section{Introduction}

Shrinkage priors are indispensable in modern Bayesian inference and  allow one to address
model specification uncertainty in a principled manner. One particularly relevant area of application,
with a rich variety of potentially useful shrinkage priors,
 is  Bayesian factor analysis.

\maketitle

In factor analysis it is assumed that the covariance matrix
$\Vary _0=\betav   _0 \betav  _0^ \top + \Vare  _0$ of $n$
multivariate observations $\ym_t=(y_{1t}, \ldots,y_{mt})^ \top$, $t=1,\ldots,n$, of dimension $m$
is generated from the
Gaussian factor model
\begin{equation}\label{fac1}
 \ym_t=  \betav _0 \facm_t   + \errorm_t,
\end{equation}
 where $\errorm_t \sim \Normult{m}{\bfz,
\Vare _0}$ are idiosyncratic errors with
 $\Vare  _0=\Diag{\sigma_1^2, \ldots, \sigma_m^2}$ and $\facm_t
\sim \Normult{\nfactrue}{\bfz,\identm}$ are latent factors of factor dimension $\nfactrue$.

In applied factor analysis, the dimension $\nfactrue$
of the factor space is typically not known and has to
be inferred from the data.
The Bayesian approach provides an attractive solution to this problem,
since the unknown factor dimension $\nfactrue$ can be estimated 
in an overfitting factor model
along with all other unknown parameters,
such as the factor loadings $\{\beta_{ih}\}$ in the loading matrix $\betav _0 \in \Real^{m \times \nfactrue}$
and
the idiosyncratic variances $\sigma_1^2, \ldots, \sigma_m^2$.
In sparse Bayesian factor analysis, the strategy to
recover the number of factors   relies on inducing
 zero columns in the loading matrix of a factor model where $H > \nfactrue $ columns are assumed, purposefully overfitting the true, but unknown  factor dimension
 $\nfactrue$. In finite factor analysis,  $H \leq (m-1)/2$ is chosen to ensure econometric identification (see for instance \cite{fru-etal:whe} and \cite{hos-fru:cov}),
 whereas in infinite factor analysis $ H=\infty$; see  \cite{bha-dun:spa} for pioneering work in this area.
  Shrinkage priors are then placed on the factor loadings, with the goal of automatically removing
   all redundant
  columns based on the information in the data.
 There are basically four main approaches for choosing shrinkage priors
in sparse Bayesian factor analysis.

One strand of literature works with continuous shrinkage priors in finite factor analysis,
often in the context of efficient
estimation of the covariance matrix $\Vary_0$, see e.g. \cite{kas:spa}. While these priors implicitly reduce
the dimension of the parameter space, it is not straightforward how to explicitly retrieve the
unknown factor dimension $\nfactrue$.

In infinite factor analysis, \cite{bha-dun:spa} also work with continuous shrinkage priors on the
factor loadings.
They introduce the multiplicative gamma process (MGP) prior with the aim to penalize the effect of
additional
columns in the factor loading matrix. The MGP prior defines the prior precision of all factor loadings in
a specific column as a cumulative product of gamma priors. This prior has been widely applied,
 e.g. by \cite{mur-etal:inf} in the context of infinite mixtures of factor analyzers
 and by \cite{dev-etal:bay} in the context of Bayesian multi-study factor analysis for
 high-throughput biological data. However,
 \cite{dur:not} shows that the intended goal of increasing shrinkage is achieved
 only for specific settings
 of hyperparameters. As a result, the method tends to overestimate the true number of factors,
 as demonstrated by \cite{leg-etal:bay} in a comprehensive simulation study.

 A third, extremely rich  strand of literature  works with exchangeable 
 spike-and-slab  priors with 
 column-specific probabilities assigned to the spike and to the slab,
 see  \cite{fru-etal:spa,wes:bay_fac,car-etal:hig,teh-etal:sti,fru-lop:par,con-etal:bay_exp,roc-geo:fas,kau-sch:bay},
 among many others.
 More specifically,
a binary indicator  $\delta_{ih}$ is introduced for each element $\beta_{ih}$  of the loading matrix
and  a column-specific occurrence
probability $\Prob{\delta_{ih}=1|\tau_{h}}=\tau_{h}$
 for non-zero elements in each column $h$ of the factor loading matrix
is assumed.
 As opposed to the MGP prior,
 an exchangeable  prior for the slab probabilities
 $\tau_{1}, \ldots, \tau_H$ across all columns is employed and no explicit  prior
ordering  or increasing shrinkage is imposed on the columns of the loading matrix.
  A popular example of such an exchangeable prior in finite
  Bayesian factor analysis is the
  one parameter beta prior
$\tau_h| H \sim  \Betadis{\frac{\alpha}{H},1}$
\cite{roc-geo:fas,fru-etal:spa,ava-etal:het}.
 As opposed to continuous shrinkage priors, the discrete nature of spike-and-slab priors allows
  explicit inference with regard to the unknown factor dimension $\nfactrue$.

 Finally, as an alternative to any of these priors, \cite{leg-etal:bay} recently
 introduced the cumulative shrinkage
 process   (CUSP) prior. In the context infinite factor models, the CUSP prior is a spike-and-slab prior where the columns of the loading matrix
 are ordered and an
 increasing prior probability is assigned to the spike as the column index increases.
 The CUSP prior is designed to capture the expectation that additional columns in the loading matrix
 will play a progressively less important role and the associated parameters  have a
 stochastically decreasing effect. In constructing the CUSP prior, \cite{leg-etal:bay} exploit
 the stick-breaking representation of  a Dirichlet process (DP) prior \cite{set:con}.
 Recently,
  \cite{kow-can:sem} extended the CUSP prior in two ways, first by considering a more general spike
  distribution and, second, by using the stick-breaking representation of  the two-parameter Indian buffet process
  prior introduced by \cite{teh-etal:sti}. The authors apply this
  \lq\lq ordered spike-and-slab prior\rq\rq\
  in the context of semi-parametric functional factor models.

The present paper makes two  main contributions in this research field.  First, the
cumulative shrinkage
 process  priors  of
\cite{leg-etal:bay} and \cite{kow-can:sem} are extended
to the class of generalized cumulative shrinkage
 process   priors,
by involving  very general stick-breaking representations which might be finite or infinite.
It is proven that the ordering in the spike probabilities induces increasing shrinkage
for the parameters of interest, as has been proven in
\cite[Proposition~1]{kow-can:sem}  for ordered spike-and-slab  priors (which contain
the CUSP  prior of \cite{leg-etal:bay} as a special case).
The generalized CUSP prior subsumes several specific priors involving stick-breaking representations
from beta distributions that were
introduced earlier
 in the literature for factor-analytical models, see e.g.
\cite{roc-geo:fas,hea-roy:gib,ohn-kim:pos,fru-etal:spa,kow-can:sem}.

Second, we shed new light on the  popular class of exchangeable shrinkage process
 priors, including one and two parameter beta priors. We show that  any exchangeable spike-and-slab prior on a sequence of parameters
 has a representation as a generalized cumulative shrinkage
 process
 prior and implicitly  imposes increasing shrinkage on the parameters.
This representation can be simply derived from the decreasing order statistics of the slab probabilities.
Finally, we discuss applications of this generalized CUSP prior in the context of finite
sparse  Bayesian factor models.

The rest of the paper is organized as follows. Section~\ref{sec:CUSP}
introduces the generalized cumulative shrinkage
 process prior and provides several examples.  Section~\ref{sec:EPS} shows how exchangeable shrinkage process
  priors can be expressed as generalized CUSP priors.  Section~\ref{parc} discusses posterior inference for both classes of priors.
 Section~\ref{sec:app} illustrates applications to sparse Bayesian factor analysis
  and Section~\ref{final} concludes.

\section{Generalized cumulative shrinkage  process priors}         \label{sec:CUSP}

\subsection{Definition} \label{defCUSP}

Cumulative shrinkage
 process  (CUSP) priors were introduced by
\cite{leg-etal:bay} to induce increasing shrinkage on a countable sequence of model parameters
$\{\theta_h\}, h=1, \ldots, H$.
Increasing shrinkage is achieved  by
assigning a spike-and-slab prior to each parameter $\theta_h$,
 \begin{eqnarray} \label{CSPleg}
 \theta_h| \piL_h \sim  \piL_h \diracarg{\theta_\infty} + (1- \piL_h) P_{\tiny \rm slab} (\theta_h), 
 \end{eqnarray}
 where an
 increasing prior probability $\piL_h $ is assigned to a Dirac spike at  $\theta_\infty$.
 Based on the stick-breaking representation $ \stick_h \,\, i.i.d. \, \, \Betadis{1,\alpha}$
  of a Dirichlet process (DP) \cite{set:con}, the sequence of increasing spike
  probabilities $\piL_h $ is defined as:
  \begin{eqnarray} \label{CSP2leg}
 \piL_h= \sum_{\ell=1}^h \omega_\ell, \quad
 \omega_\ell=\stick_\ell \prod _{j =1}^{\ell-1}(1- \stick_j).
 \end{eqnarray}
 Evidently, the spike probabilities in (\ref{CSP2leg}) are increasing, since
$\piL_{h}=\piL_{h-1} + \omega_{h}$ with $\omega_{h} \in (0,1)$.
For $H=\infty$, the sequence $\{\omega_h\}$ defined
 in (\ref{CSP2leg}) is the stick-breaking representation of the  weights $\{\omega_h\}$ of a DP mixture. Hence, 
 $\sum_{\ell=1}^\infty \omega_\ell=1$ and $\piL_h$ approaches 1 as $h$ increases.
  A finite (truncated) version of the CUSP prior is obtained by
   defining $ \stick_H=1$ for some finite $H< \infty$.

 Recently, \cite{kow-can:sem} introduced ordered spike-and-slab  priors
 which  generalize the CUSP prior defined in (\ref{CSPleg}) in various directions.
 First, the authors consider general spike
  distributions, 
   \begin{eqnarray} \label{CSP1KC}
 \theta_h| \piL_h \sim  \piL_h P_{\tiny \rm spike} (\theta_h)
  + (1- \piL_h) P_{\tiny \rm slab} (\theta_h),
 \end{eqnarray}
 while \cite{leg-etal:bay} assume
 a Dirac spike $\diracarg{\theta_\infty}$ at a
  known small value $\theta_\infty$. However, as shown by \cite{sch-can:tru}
  in the context of infinite factor models, the
  choice of this parameter can be very influential.

  Second, \cite{kow-can:sem}
  construct  the sequence of increasing spike probabilities
  $\{\piL_h\}, h=1, \ldots, \infty $
   in (\ref{CSP2leg}) in a more general manner, namely as a cumulative process
 involving
  the stick-breaking representation
   $ \stick_h \,\, i.i.d. \, \, \Betadis{\betaTau, \betaTau \alpha}$
  of  the two-parameter Indian buffet process (IBP)
  prior introduced by \cite{teh-etal:sti}.
   With $\betaTau=1$,  the stick-breaking representation $ \stick_h \,\, i.i.d. \, \,
  \Betadis{1, \alpha}$ results and  the ordered spike-and-slab prior reduces to the CUSP prior
  of \cite{leg-etal:bay}.

 The strength parameter  $\alpha$ in the CUSP prior plays an important role
  in determining  how many parameters $\theta_h$
   are active and is assumed to be known and fixed  at $\alpha=5$  in \cite{leg-etal:bay}. In contrast to this,
  \cite{kow-can:sem} allow
$\alpha$ (called $\kappa$ in their paper)
to be an unknown hyperparameter that is learned from the data
under a gamma prior, $\alpha \sim \Gammad{\aalpha, \balpha}$,
while $\betaTau$ (called $\iota$ in their paper) is fixed and typically chosen as $\betaTau=1$.

The present paper extends this important work further and
introduces a generalized CUSP prior in Definition~\ref{def31}. Specifically, the sequence of increasing spike probabilities
  $\{\piL_h\}, h=1, 2, \ldots$
   is constructed  as a cumulative process
 involving more general (and possibly finite) stick-breaking constructions
 $ \{\stick_h\}$, which need not arise
 from the same distribution.
 The CUSP priors  introduced by \cite{leg-etal:bay} and \cite{kow-can:sem}
 are special
 cases of this  generalized CUSP prior. Additional examples are discussed in 
\citeSec{sec:CUSP}{sec:example} and 
\citeSec{sec:EPS}{sec:ex:ESP}.

 \begin{definition}[Generalized cumulative shrinkage process prior] \label{def31}
 Let $\piLv=\{\piL_h \in (0,1)\}, h=1, \ldots, H$ be a countable sequence of random
 parameters
 taking values in the unit interval which are defined by:
 \begin{eqnarray} \label{CSP2}
 \piL_h= \sum_{\ell=1}^h \omega_\ell, \quad  \omega_\ell=\stick_\ell
  \prod _{j =1}^{\ell-1}(1- \stick_j),
 \end{eqnarray}
 where $ \{\stick_h\} $,  $h=1, \ldots, H$ is a sequence of random variables
 taking values in the unit interval.
  For $H< \infty$, 
  $\stick_H$ can, but need not take the value 1.
   For $H = \infty$,
  it is assumed that $\sum_{\ell=1}^\infty \omega_\ell=1$ almost surely.
  Let $\Theta=\{\theta_h\}, h=1, \ldots, H$ be a countable sequence of model parameters.
Assume that the  parameters $\theta_h$ are independent conditional
on $\piLv$ and that $ \theta_h| \piL_h$ is independent of $\piL_\ell, \ell \neq h$ for all $h$.
Assume that $p (\theta_h| \piL_h)$ takes the form of following
  spike-and-slab prior:
  \begin{eqnarray} \label{CSP1}
 \theta_h| \piL_h \sim  \piL_h P_{\tiny \rm spike} (\theta_h)
  + (1- \piL_h) P_{\tiny \rm slab} (\theta_h).
 \end{eqnarray}
 Then, for $H< \infty$, $\Theta$ is said to follow a finite
 generalized cumulative shrinkage process (CUSP) prior. If $H = \infty$,
 then $\Theta$ is said to follow an infinite
 generalized CUSP prior.
 \end{definition}
\noindent Note that the spike probabilities $\piL_{h}=\piL_{h-1} + \omega_{h}$ in definition (\ref{CSP2}) are an increasing sequence by construction and
$ \Ew{\piL_{h}} >  \Ew{\piL_{h-1}}$.
The ordering of the spike probabilities $\{\piL_h\}$ in the generalized CUSP prior  implies an explicit ordering for the prior
 distributions of the parameters  $\{\theta_h\}$   in (\ref{CSP1}).
 This has been proven in \cite[Proposition~1]{kow-can:sem} for the ordered spike-and-slab
  prior, extending \cite[Lemma~1]{leg-etal:bay}.
  It follows from a straightforward extension of the
  corresponding proof that this important property also holds for the
 generalized CUSP prior introduced in Definition~\ref{def31}. This insight is summarized
 in  Proposition~\ref{prop41}.

    \begin{proposition} \label{prop41}
    For $\varepsilon>0$ and a fixed $\theta_0$, let
     $\mathbb{B}_\varepsilon (\theta_0)=\{ \theta_h: |\theta_h - \theta_0| < \varepsilon \}$.
      Under prior (\ref{CSP2}), whenever
    the spike and the slab distribution in (\ref{CSP1}) satisfy
  \begin{eqnarray} \label{decsp}
 P_{\tiny \rm spike} (\mathbb{B}_\varepsilon (\theta_0)) >  P_{\tiny \rm slab} (\mathbb{B}_\varepsilon (\theta_0)),
  \end{eqnarray}
  then
 \begin{eqnarray} \label{prop4Un}
 \Prob{|\theta_h - \theta_0| \leq \varepsilon} <   \Prob{|\theta_{h+1} - \theta_0| \leq \varepsilon}.
  \end{eqnarray}
  \end{proposition}
The proof is a straightforward extension of the corresponding proof
of \cite[Proposition~1]{kow-can:sem} to generalized CUSP priors.
For any $\varepsilon>0$ and  fixed $\theta_0$, the following holds:
\begin{eqnarray*}
   \Prob{|\theta_h - \theta_0| \leq \varepsilon} 
&=&  P_{\tiny \rm slab} (\mathbb{B}_\varepsilon (\theta_0)) \Ew{1- \piL_h} +
    P_{\tiny \rm spike} ( \mathbb{B}_\varepsilon (\theta_0) ) \Ew{\piL_h} \\
   &=& P_{\tiny \rm spike} ( \mathbb{B}_\varepsilon (\theta_0) )  + (1-\Ew{\piL_h})
   \left( P_{\tiny \rm slab} (\mathbb{B}_\varepsilon (\theta_0)) - P_{\tiny \rm spike} ( \mathbb{B}_\varepsilon (\theta_0) )\right).
\end{eqnarray*}
Since $\Ew{\piL_h}$ is strictly increasing in $h$,
(\ref{prop4Un}) follows immediately (provided that (\ref{decsp}) holds):
\begin{eqnarray*}
\Prob{|\theta_{h+1} - \theta_0|\leq \varepsilon} - \Prob{|\theta_h - \theta_0| \leq \varepsilon}=
\left( \Ew{\piL_{h+1}} -\Ew{\piL_{h}} \right)
   \left(  P_{\tiny \rm spike} ( \mathbb{B}_\varepsilon (\theta_0) ) - P_{\tiny \rm slab} (\mathbb{B}_\varepsilon (\theta_0)) \right) >0 .
\end{eqnarray*}
It is also interesting to verify that
    the decreasing  sequence  of slab probabilities $\piL^\star_h= 1 - \piL_h$
   has following representation:
     \begin{eqnarray} \label{CSP3}
 \piL^\star_h= \prod _{\ell =1}^{h} (1-\stick_\ell)=
  \prod _{\ell =1}^{h} \nu^\star_\ell,   \qquad h=1,2, \ldots, H.
  \end{eqnarray}
  This  result is easily proven by induction.
 (\ref{CSP3}) obviously holds for $h=1$, since $\piL^\star_1= 1 - \piL_1 = 1- \omega_1= 1- \stick_1=
 \nu^\star_1$.
 Assume that (\ref{CSP3}) holds up to $h-1$. Then $ \piL_{h}= \piL_{h-1} + \omega_{h}$, where
 \begin{eqnarray*} \omega_{h}=\stick_{h} \prod _{\ell =1}^{h-1} (1- \stick_\ell)=
\stick_{h} \piL^\star_{h-1},
 \end{eqnarray*}
 and we obtain:
    \begin{eqnarray*} \piL^\star_{h}= 1-\piL_{h} = 1-  \piL_{h-1} - \omega_{h}=
 \piL^\star_{h-1} - \stick_{h} \piL^\star_{h-1} = (1 - \stick_{h})
 \piL^\star_{h-1} = \prod _{\ell =1}^{h} \nu^\star_{\ell}.
 \end{eqnarray*}

\subsection{Examples of CUSP priors} \label{sec:example}

Definition~\ref{def31} is rather generic and does not make any specific assumptions
regarding the  sequence of random variables
$ \{\stick_h\} $,  $h=1, \ldots, H$. In Section~\ref{sec:EPS}, we will
show how  $ \{\stick_h\} $ can be derived from the decreasing order statistics of the
slab probabilities in a finite exchangeable shrinkage process prior.

Alternatively, the sticks $ \{\stick_h\} $ can be chosen to come from  a specific distribution family. For example, they could arise
as independent random variables from beta distributions:
 \begin{eqnarray} \label{CSP2ext}
 \stick_h  \sim \Betadis{a_h,b_h}, \quad h=1, \ldots, H.
 \end{eqnarray}
 Exploiting  (\ref{CSP3}), the decreasing slab probabilities $\piL^\star_h$ can be presented as
a multiplicative beta process  with
  $\nu^\star_h \sim \Betadis{b_h,a_h}$.

  Several special cases of such a  shrinkage prior have been suggested in the literature.
  Obviously, the CUSP prior introduced by \cite{leg-etal:bay} results as a special
 case of (\ref{CSP2ext}), where $H=\infty$, $a_h=1$ and $b_h=\alpha$.
 As noted by \cite{teh-etal:sti}, this prior is equivalent to the IBP prior.
\cite{gha-etal:bay}  define 
 the two-parameter Indian buffet process
  prior  from  the stick-breaking representation
   $ \stick_h \,\, i.i.d. \, \, \Betadis{\betaTau, \betaTau \alpha}$, extending the IBP prior of \cite{teh-etal:sti}.
   The  ordered spike-and-slab prior
 of \cite{kow-can:sem}  is based on this   stick-breaking representation and
 results as a special
 case of (\ref{CSP2ext})  where $H=\infty$,
  $a_h=\betaTau$ and $b_h=\betaTau \alpha$.
In the context of high-dimensional sparse factor models,
  \cite{ohn-kim:pos} define the sequence of slab probabilities $ \piL^\star_{h}$
  in a spike-and-slab prior for the factor loadings
  as in
  (\ref{CSP3})  with
  $\nu^\star_h \sim \Betadis{\alpha,1+\kappa}$, where $\alpha >0$ and $\kappa \geq 0$
  are hyperparameters. This prior
  results as a special
 case of (\ref{CSP2ext})  where $H=\infty$,
  $a_h=1+\kappa$ and $b_h=\alpha$. Further, it leads to a two-parameter IBP prior with an alternative
  parameterization
  vis-\`a-vis the prior applied in \cite{kow-can:sem}.

Another way to construct generalized CUSP priors is to exploit the weights of more general
mixtures than DP mixtures. In principle, the weights of any finite or infinite mixture
can be used to define a generalized CUSP prior, 
see \cite[Figure~1]{teh-etal:sti}.
Examples include
 the Pitman-Yor-Process (PYP)-prior which has been applied by  \cite{hea-roy:gib}
 to define Gibbs-type Indian buffet processes. Choosing
 $\stick_h \sim \Betadis{1 - \sigma ,\alpha + h \sigma}$ with
  $\sigma \in [0,1)$ and $\alpha > \sigma$ in (\ref{CSP2ext})
implies a large number of active coefficients with significant, but small weights $\piL^*_h$.
For $\sigma=0$, the induced generalized CUSP prior reduces to the CUSP prior of  \cite{leg-etal:bay}.
 Alternatively, one could choose
 the PYP-prior, 
 $\stick_h \sim \Betadis{1 - \sigma , (H - h) |\sigma|}$, $h=1, \ldots,H-1$,
 where $\sigma <0$ is negative and  $H$ is a natural
 number. Choosing  this stick-breaking representation in (\ref{CSP2ext})
leads to a generalized CUSP
with a finite number $H$ of active coefficients, where $a_h=1 - \sigma$
 and $b_h=(H - h) |\sigma|$.

\section{Exchangeable shrinkage process priors} \label{sec:EPS}

\subsection{Definition} \label{defESP}

  \begin{definition}[Exchangeable shrinkage process priors] \label{def31EPS}
 Let $\tauv=\{\tau_h \in (0,1)\}, h=1, \ldots, H$ with $H<\infty$ be a finite sequence of
 iid random parameters taking values in the unit interval.
 Let $\Theta=\{\theta_h\}, h=1, \ldots, H$ be a finite sequence of model parameters
 and assume that the parameters $ \theta_h| \tau_h$ are independent conditional
on $\tauv$ and  independent of $\tau_\ell, \ell\neq h$ for all $h$.
If  $p (\theta_h| \tau_h)$ takes the form of a spike-and-slab prior:
  \begin{eqnarray} \label{CSP1E}
 \theta_h| \tau_h \sim  (1-\tau_h) P_{\tiny \rm spike} (\theta_h) +  \tau_h  P_{\tiny \rm slab} (\theta_h),
 \end{eqnarray}
 then  $\Theta$ is said to follow an exchangeable shrinkage process (ESP)
 prior.
  \end{definition}
  By definition, prior  (\ref{CSP1E})  is invariant to permuting the indices of
$\theta_h$. Hence, if a sequence $\{\theta_h\}, h=1, \ldots, H$ implies
an exchangeable shrinkage process (ESP) prior, then for any permutation
 $\rho(1), \ldots,  \rho(H)$ of the indices $1, \ldots, H$, the sequence
$\{\theta_{\rho(h)}\}, h=1, \ldots, H$ follows the same ESP prior.

To complete the definition of an ESP prior, a probability law for the slab probabilities
$\tau_1, \ldots, \tau_H$
has to be chosen. Typically, it is assumed that
\begin{eqnarray} \label{prigen}
\tau_h| H \sim \Betadis{a_0,b_0}, \qquad h=1,\ldots, H,
 \end{eqnarray}
with  $a_0$ and $b_0$ potentially depending on $H$ as well as on unknown hyperparameters.

\paragraph*{Representation as a CUSP prior.}

A main contribution of this paper is to prove that any  ESP prior admits a
 finite generalized CUSP representation as in (\ref{CSP2}) and (\ref{CSP1}).
The  CUSP representation  is  obtained
 by a simple permutation of the indices $1, \ldots, H$. Consider the
   decreasing order statistics
   $\tau_{(1)} > \ldots > \tau_{(H)}$ of the unordered slab probabilities
   $\tau_1, \ldots, \tau_H$ of prior (\ref{CSP1E}). If we permute the coefficients
   $\theta_1, \ldots, \theta_H$
    according to
    the decreasing slab probabilities $\tau_{(1)}, \ldots, \tau_{(H)}$,
    then the spike probabilities $\piL_h$
   in the generalized CUSP representation
    are equal to
    $\piL_h= 1 - \tau_{(h)}$ for $h=1, \ldots, H$ and are increasing by definition.
   Hence, by the virtue of
Proposition~\ref{prop41},
an ESP prior induces increasing shrinkage
for the sequence of  ordered coefficients $\theta_{\rho(1)}, \ldots, \theta_{\rho(H)}$,
where the parameters $\theta_1, \ldots, \theta_H$ are ordered according to the permutation
underlying the decreasing order statistics $\tau_{(1)}, \ldots, \tau_{(H)}$.
Therefore, increasing shrinkage is achieved  without
  explicitly imposing any  ordering on the spike probabilities in the definition of the ESP prior.

It should be emphasized that we do not need to know the explicit CUSP representation to
achieve this shrinkage property.
 Theoretically,  we could derive the distribution of the sticks $\nu^\star_h$ or, equivalently,
  $\stick_h$  from (\ref{CSP3}) based on the decreasing order statistics $\tau_{(h)} < \tau_{(h-1)}$:
    \begin{eqnarray*}
    \nu^\star_h =  \frac{\tau_{(h)}}{\tau_{(h-1)}}, \qquad
     \stick_h =  1- \frac{\tau_{(h)}}{\tau_{(h-1)}}.
    \end{eqnarray*}
   However, only in specific cases  will it be possible to work out
    the explicit distribution of the sticks $\{\stick_h\}, h=1, \ldots, H$, see
\citeSec{sec:EPS}{sec:ex:ESP}
   for an example.
In any case, $\piL_h= 1 - \tau_{(h)}$ is an increasing sequence, such that
    $\Ew{\piL_{h+1}} >  \Ew{\piL_{h}}$.
    This is all we need to prove Proposition~\ref{prop41} for the sequence of
    ordered coefficients $\theta_{\rho(1)}, \ldots, \theta_{\rho(H)}$.

 \subsection{Examples of ESP priors} \label{sec:ex:ESP}

 Exchangeable shrinkage process priors have been applied by many authors,
  in particular in sparse Bayesian factor analysis.
  \cite{fru-etal:spa}, for instance, assume that the hyperparameter $a_0$ in
 (\ref{prigen}) is  dependent on $H$ by choosing $a_0=\frac{\alpha}{H}\betaTau$ and $b_0=\betaTau$:
 \begin{eqnarray} \label{pri2Pgen}
 \tau_h| H \sim  \Betadis{\frac{\alpha}{H}\betaTau,\betaTau},
 \end{eqnarray}
 where $H$ is a maximum number of potential factors and $\alpha$ and $\betaTau$ are hyperparameters that can be estimated from the data.
 For $H \rightarrow \infty$, the finite two-parameter beta (2PB)
 prior (\ref{pri2Pgen}) converges to the infinite 2PB
  prior  introduced by \cite{gha-etal:bay}
 in the context of Bayesian nonparametric latent feature models.  These can be regarded as a
  factor model with infinitely many columns of which only a finite number is non-zero.

For $\betaTau=1$, the finite one parameter  beta (1PB) prior employed by \cite{roc-geo:fas}
 results:
 \begin{eqnarray} \label{prialt1P}
 \tau_h| H \sim  \Betadis{\frac{\alpha}{H},1}.
 \end{eqnarray}
 It is well-known that
 this prior converges to the Indian buffet process (IBP) prior 
  for $H \rightarrow \infty$, see \cite{teh-etal:sti}.
  In the Appendix
   it is shown
 that  the generalized CUSP representation of prior  (\ref{prialt1P}) involves
 the stick-breaking representation
  $\stick_h   \sim  \Betadis{1,\alpha \frac{H-h+1}{H}}$, making it a special case of
 the generalized CUSP prior (\ref{CSP2ext}). Therefore, as $H$ goes to infinity, the  finite 1PB  prior
 (\ref{prialt1P}) converges to the
 CUSP prior proposed by \cite{leg-etal:bay} with
 strength parameter $\alpha$.

\section{Posterior inference} \label{parc}

Posterior inference for both the ESP prior
 as well as the general CUSP prior is based on Markov chain Monte Carlo (MCMC) estimation, with data augmentation proving to be particularly useful.

Depending on the application context, $\theta_h$ typically acts as a hyperparameter for a hierarchical
prior $\betav_h|\theta_h$ involving additional model parameters $\betav_h$ (e.g. the column-specific factor loadings $\betav_h=(\beta_{1h}, \ldots, \beta_{mh}) ^\top$).
Marginalizing over $\theta_h$  yields
the following spike-and-slab prior for $\betav_h$:
  \begin{eqnarray} \label{Cbeta1}
 \betav_h| \piL_h \sim  \piL_h
  P_{\tiny \rm spike} (\betav_h) + (1-  \piL_h) P_{\tiny \rm slab} (\betav_h),
 \end{eqnarray}
 where the    
  pdfs of, respectively, the spike and the slab distribution are given by:
 \begin{eqnarray*} p_{\tiny \rm spike} (\betav_h) = \int p(\betav_h|\theta_h)  p_{\tiny \rm spike} (\theta_h) \, d  \theta_h, \quad
 p_{\tiny \rm slab} (\betav_h) = \int p(\betav_h|\theta_h)  p_{\tiny \rm slab} (\theta_h) \, d  \theta_h.
 \end{eqnarray*}

\subsection{Data augmentation and MCMC for ESP priors} \label{sec:DAEPS}

Data augmentation and MCMC for exchangeable shrinkage process (ESP)
priors has been considered in numerous papers.
For prior (\ref{CSP1E}), a binary indicator variable $S_h$ with Bernoulli prior $\Prob{S_h =1|\tau_h}=\tau_h$ is
introduced for each $h=1, \ldots,H$.
Given $S_h$,
the parameter $\theta_h$  is then classified a priori  into spike or slab:
\begin{eqnarray*} \theta_h| S_h \sim  (1-S_h)  P_{\tiny \rm spike} (\theta_h) + S_h  P_{\tiny \rm slab} (\theta_h).
 \end{eqnarray*}
Within an MCMC scheme,  the indicators $S_1, \ldots,
 S_H$ as well as 
 the  slab probabilities
 $\tau_{1}, \ldots, \tau_{H}$  are introduced as unknowns and sampled from the respective conditional posteriors.

  Sampling the indicator $S_h $ operates on a $H \times 2$ grid and can be implemented in various ways. In the spirit of \cite{leg-etal:bay}, classification can be performed
conditional on the parameters
$\betav_1, \ldots , \betav_H$ and the slab probabilities $\tau_{1}, \ldots, \tau_{H}$,
  \begin{eqnarray*}
 \Prob{S_h =0| \betav_h, \tau_h} \propto     (1-\tau_h)  p_{\tiny \rm spike} (\betav_h) , \quad
 \Prob{S_h =1| \betav_h, \tau_h} \propto   \tau_h  p_{\tiny \rm slab} (\betav_h) ,
  \end{eqnarray*}
  where $p_{\tiny \rm spike} (\betav_h)$ and $p_{\tiny \rm slab} (\betav_h)$ are the
  pdfs of, respectively, the spike and the slab distribution in (\ref{Cbeta1}).
A more efficient sampler is obtained by two modifications.
  First, 
  by sampling $S_h$
  marginalized w.r.t. $\tau_{1}, \ldots, \tau_{H}$. 
  Second, instead of the  multivariate mixture (\ref{Cbeta1}) 
  which becomes rather informative as the dimension of $\betav_h$ increases,
  the mixture prior (\ref{CSP1E}) on $\theta_h$ can be  
  exploited for classification based 
  directly on  $\theta_1, \ldots, \theta_H$. These modifications yield:
\begin{eqnarray}  \label{sec4a1}
\Prob{S_h =0| \theta_h, q_A} \propto  (1- q_A) \cdot 
p_{\tiny \rm spike} (\theta_h), \quad 
\Prob{S_h =1| \theta_h, q_A} \propto  q_A \cdot 
 p_{\tiny \rm slab} (\theta_h) ,
\end{eqnarray}
where $q_A = \Ew{\tau_h}=\frac{a_0}{a_0+b_0} $ is the expected prior probability of the slab and $p_{\tiny \rm spike} (\theta_h)$ and $p_{\tiny \rm slab} (\theta_h)$ are the  pdfs of, respectively, the spike and the slab distribution in (\ref{CSP1E}).

 In any case, the slab probabilities
 $\tau_{1}, \ldots, \tau_{H}$ are then updated conditional on the
  indicators $S_1, \ldots,
 S_H$, by sampling $\tau_{h}$ from $\tau_{h} | S_h$
  for $h=1, \ldots, H$. Under the prior (\ref{prigen}), this yields
  \begin{eqnarray} \label{sametau}
  \tau_{h} | S_h \sim  \Betadis{a_0+ S_h , b_0 +  1 - S_h}.
  \end{eqnarray}

\subsection{Data augmentation and MCMC for CUSP priors} \label{sec:DACUSP}

 To perform MCMC for the  CUSP prior,
  \cite{leg-etal:bay} truncate the 
  infinite representation (\ref{CSP2leg}) at $H< \infty$
  and introduce
  $h$ categorical indicators $z_1,  \ldots, z_H$. Each $z_h$ takes values
  in $\{1,2, \ldots, H\}$ with the discrete prior distribution
  $\Prob{z_h=\ell}=\omega_\ell$,
    $\ell=1, \ldots,H$.
        Given $z_h$, the
 spike-and-slab prior (\ref{CSP1}) is represented as:
  \begin{eqnarray} \label{CSPDA1}
 \theta_h| z_h \sim  \indic{z_h \leq h} P_{\tiny \rm spike} (\theta_h)
  + (1-  \indic{z_h \leq h}) P_{\tiny \rm slab} (\theta_h).
 \end{eqnarray}
This data augmentation technique is generic and can be applied
to the generalized CUSP prior introduced in Definition~\ref{def31} without any modification.

 In addition, \cite{leg-etal:bay} introduce
 the sticks $\stick_{1}, \ldots, \stick_{H}$  as unknowns, which
  are sampled from their respective conditional posteriors given the categorical indicators
   $z_1, \ldots, z_H$. This step is easily extended to a generalized CUSP prior induced by
    a stick-breaking presentation $ \stick_{\ell}\sim \Betadis{a_\ell,b_\ell}$ arising from the beta distribution,
    see also \cite{kow-can:sem}.
The sticks $\stick_{1}, \ldots, \stick_{H}$ are updated conditional on the
  indicators $z_1, \ldots, z_H$, by sampling $ \stick_{\ell}$ from $ \stick_{\ell}| z_1, \ldots, z_H$
  for $\ell=1, \ldots, H$:
  \begin{eqnarray*}
  \stick_{\ell}| z_1, \ldots, z_H \sim \Betadis{a_\ell + \sum_{h=1}^H \indic{z_h =\ell}  ,b_\ell +
   \sum_{h=1}^H \indic{z_h >\ell }}.
  \end{eqnarray*}
      For $(a_\ell,b_\ell)=(1,\alpha)$ and $(a_\ell,b_\ell)=(\betaTau,\betaTau\alpha)$,  respectively,  the   sampling steps in \cite[Algorithm~1]{leg-etal:bay} and \cite[Algorithm~1]{kow-can:sem} result.  Given the sticks $\stick_{1}, \ldots, \stick_{H}$,
    the weights $\omega_1, \ldots, \omega_H$
    and the spike probabilities $\piL_1, \ldots, \piL_H$
     are updated based on (\ref{CSP2}).

Sampling the categorical indicators $z_1, \ldots, z_H$ operates on an $H \times H$ grid,
  conditional on the parameters
$\betav_1, \ldots , \betav_H$ and the  weights $\omega_1, \ldots, \omega_H$:
 \begin{eqnarray*}
 \Prob{z_h =\ell| \betav_h} \propto
 \left\{
   \begin{array}{ll}
     \omega_\ell \, p_{\tiny \rm spike} (\betav_h) , & \ell=1, \ldots, h, \\[2mm]
     \omega_\ell \, p_{\tiny \rm slab} (\betav_h), & \ell=h+1, \ldots, H. \\
   \end{array}
 \right.
  \end{eqnarray*}
Given the indicators $z_1, \ldots, z_H$, the coefficients
$\theta_1, \ldots , \theta_H$ are sampled, respectively, from the spike or the slab,
using the representation $\theta_h|z_h$ given in (\ref{CSPDA1}).

 The CUSP prior also admits a representation involving binary indicator variables
 $S _1, \ldots,  S _H$ which are defined as
 $S _h = \indic{z_h > h}$ with prior probability
 $\Prob{S _h=1|\piL_h}=\Prob{z_h > h|\piL_h} = 1-\piL_{h}=\piL ^*_{h}$, where $\piL ^*_{h}$
is the slab probability. 
 Given $S _h$, prior (\ref{CSPDA1}) can be rewritten as:
\begin{eqnarray*} \theta_h| S _h  \sim (1-S _h) P_{\tiny \rm spike} (\theta_h)
 + S _h P_{\tiny \rm slab} (\theta_h).
 \end{eqnarray*}
One may be tempted to think that, as for ESP priors, the categorical variables $z_1, \ldots, z_H$ in the MCMC scheme for CUSP priors could be substituted by  
binary indicators $ S _1, \ldots, S _H$. This would simplify  sampling considerably. 
 However, while $S _1, \ldots, S _H$ could be sampled in a similar
manner as in 
\citeSec{parc}{sec:DAEPS},
it is not possible  to sample the sticks  based on  the binary
indicators $ S _1, \ldots, S _H$, because the prior $p(S _h|\piL ^*_{h})$ only carries
the information about the events $\{z_h > h\}$ and $\{z_h \leq h\}$, but not about $\{z_h = \ell\}$ and $\{z_h > \ell\}$ for $\ell \neq h$.

Nevertheless,  computational gains can be achieved by using a finite exchangeable
shrinkage process prior, in particular if MCMC estimation
is based on a truncated version of an
infinite CUSP prior, see
\citeSec{sec:app}{illapp} for illustration.
In this case, any finite  exchangeable
shrinkage process prior that converges to the infinite CUSP prior can be employed.
Examples are the finite 1PB prior (\ref{prialt1P}), which converges to the CUSP prior of
\cite{leg-etal:bay} and the finite 2PB prior (\ref{pri2Pgen}), which converges to the
ordered spike-and-slab prior of \cite{kow-can:sem}.

\subsection{Inference on the number of active coefficients} \label{hstar}

One of the main reasons for introducing  either an ESP  or a CUSP prior on a sequence of coefficients
is to learn how many coefficients are active.
  This procedure is used
  in \cite{leg-etal:bay} to estimate the number of active factors
 in sparse Bayesian factor analysis under the  CUSP prior (\ref{CSPleg}) 
 and is applied in
\cite{kow-can:sem} to estimate the number of active terms for Bayesian semi-parametric
 functional factor models with Bayesian rank selection under the
  ordered spike-and-slab prior (\ref{CSP1KC}).

Based, respectively, on the categorical variables $z_1, \ldots, z_H$ or the
binary  indicators $S_1, \ldots, S_{H}$,
 the number of active coefficients  $H^\star$ is defined
as:
 \begin{eqnarray} \label{CSPDA2}
 H^\star = \sum_{h=1}^H \indic{z_h > h}, \qquad H^\star  = \sum_{h=1}^H S_h.
 \end{eqnarray}
 Representation (\ref{CSPDA2}) 
  is useful to investigate 
  how the choice of
 hyperparameters impacts the prior distribution of $ H^\star$.
As shown in \cite{leg-etal:bay}, the strength parameter
 $\alpha$ strongly  influences the prior on the model dimension $H^\star$ under a CUSP prior, with both the mean $\Ew{H^\star|\alpha}$ and the variance $\V{H^\star|\alpha}$ being equal to $\alpha$.
For this reason, \cite{kow-can:sem} recommend that the hyperparameters of a CUSP prior should be learned from the data.  

Similarly, increasing shrinkage under finite ESP priors can only be achieved through suitable choices of hyperparameters.
For the finite 1PB prior (\ref{prialt1P}) with $H< \infty$, for instance,  
both the mean and the variance of $H^\star$ 
strongly depend on $\alpha$:
 \begin{eqnarray*} \Ew{H^\star |\alpha}= \frac{\alpha}{1+\alpha/H}, \qquad  \V{H^\star |\alpha}=
  \frac{\alpha}{(1+\alpha/H)^2}.
 \end{eqnarray*}
The influence of $\alpha$ becomes even more apparent when we consider the CUSP representation of the finite 1PB  prior based on the decreasing order statistics 
 $\tau_{(1)} > \ldots > \tau_{(H)}$ of the slab probabilities. 
 The  largest slab probability  $\tau_{(1)}$ follows a $\Betadis{\alpha,1} $ distribution, while 
 the subsequent slab probabilities
 $\tau_{(h)}=\tau_{(h-1)}\nu_h, \, \nu_h \sim \mathcal {B}(\alpha ((H-h+1)/H),1),$ are increasingly pulled toward zero as $h$ increases, with final sticks $\nu _{H-1 } \sim \mathcal {B} ( 2\alpha /H, 1)$ and $\nu _{H} \sim \mathcal {B} ( \alpha /H, 1)$. Hence, a prior with  $\alpha<<H$ induces prior sparsity, since the largest spike probabilities 
 $\eta_h=1-\tau_{(h)}$ are increasingly pushed towards one. 

A common prior that does not induce prior sparsity is the uniform prior $\tau_h \sim \Uniform{0,1}$, $h=1, \ldots,H$. An example of its application can be found in e.g.~in \cite{zha-etal:bay_gro}, where it is used to introduce a Dirac spike with a column-specific fixed loading, in the same vein as \cite{leg-etal:bay} for the CUSP prior.
Formally, the uniform prior can be regarded as a special case of a 1PB prior, 
where $\alpha=H$. In this case, the 
last three sticks are distributed as $\nu_{H-2} \sim \Betadis{3,1}$,  $\nu_{H-1} \sim \Betadis{2,1}$ and $\nu_{H} \sim \Betadis{1,1}$ and,   consequently, the three largest spike probabilities are not strongly pulled towards one a priori.  
Hence, such a prior is prone to overfit the number of active coefficients, as will be confirmed in our illustrative case study in 
\citeSec{sec:app}{illapp}.

For finite ESP priors, the hyperparameters are typically assumed to be known,  
  however they can easily be assumed to be unknown parameters that are learned from the data, see e.g. \cite{fru-etal:spa}.
For both types of shrinkage priors, we discuss  
how hyperparameters are sampled during MCMC estimation under suitable priors in more detail in \citeSec{parc}{sec:hyp}.

Representation (\ref{CSPDA2})  is also useful to derive the
   posterior distribution  $p(H^\star|\ym)$ of the number of active coefficients for given data
   $\ym=(\ym_1,\ldots,\ym_n)$.
    Since the categorical  variables $z_1, \ldots, z_{H}$
    and the binary  indicators $S _1, \ldots, S_{H}$
    are sampled within an MCMC scheme,
    draws from the posterior distribution  $p(H^\star|\ym)$ of the number of active coefficients
    can be derived immediately with the help of  (\ref{CSPDA2}).

 \subsection{Learning the hyperparameters} \label{sec:hyp}

For the ordered spike-and-slab prior (\ref{CSP1KC}),
 where
   $ \stick_h \,\, i.i.d. \, \, \Betadis{\betaTau, \betaTau \alpha}$, \cite{kow-can:sem} place a gamma prior $\alpha \sim \Gammad{\aalpha, \balpha}$  on the  strength parameter
   $\alpha$, while  the other hyperparameter is fixed at $\betaTau=1$, reducing the prior to a CUSP prior with unknown strength parameter.
    In practice, $\aalpha =2$ and $\balpha=1$ is chosen,
  so that $\Ew{H^*}= \V{H^*} =2$.
For efficient MCMC estimation, \cite{kow-can:sem} truncate the CUSP prior  at $H < \infty$ 
    and  assume that $\stick_{H}=1$. Under these assumptions,   the gamma prior is conditionally conjugate to the likelihood of the sticks
  $\stick_{1}, \ldots, \stick_{H-1}$ and $\alpha|v_{1}, \ldots, v_{H-1} $ can be easily updated  from the gamma distribution
\begin{eqnarray} \label{4deq}
   \alpha| v_{1}, \ldots, v_{H-1} \sim \Gammad{\aalpha +H-1,
   \balpha - \sum_{h=1} ^{H-1}  \log (1- v_{h}) }.
   \end{eqnarray}
For finite ESP priors, 
  we found it preferable to work with the marginalized posterior (where
    the slab probabilities $\tau_{1}, \ldots, \tau_{H}$ are integrated out)  to learn the unknown hyperparameters. 
    This is easily achieved for finite ESP priors based on the  beta prior (\ref{prigen}), where  $\tau_h|H \sim \Betadis{a_0,b_0}$. Under this prior,
      the likelihood   $p(S_{1}, \ldots, S_{H}|a_0,b_0)$ is available in closed form  and depends on the indicators $S_{1}, \ldots, S_{H}$  only through  the number of active coefficients  $H^\star$ defined in (\ref{CSPDA2}):
   \begin{eqnarray*}
  p(S_{1}, \ldots, S_{H}|a_0,b_0)=
( q_A) ^{H^\star} (1-q_A) ^{H-H^\star},  \quad q_A = \Prob{S_h=1} = \frac{a_0}{a_0+b_0} .
   \end{eqnarray*}
This likelihood can be combined with a suitable prior, whenever $a_0$ and/or $b_0$
depend on unknown hyperparameters
such as $\alpha$ in the finite 1PB prior
defined  in (\ref{prialt1P}). In this case,
 $q_A=\alpha/(\alpha+H)$ and the posterior $p(\alpha| S_1, \ldots, S_H)=p(\alpha| H^\star )$ under the
 gamma prior $\alpha \sim \Gammad{\aalpha, \balpha}$  reads:
   \begin{eqnarray}  \label{AMAM}
  p(\alpha| H^\star ) \propto p(\alpha) \frac{\alpha ^{H^\star}}{(\alpha+H)^H} = \frac{\alpha ^{H^\star+ \aalpha-1}}{(\alpha+H)^H}
  \exp \left(  - \alpha \balpha \right) .
  \end{eqnarray}
  This allows for an easy implementation of an MH step to sample $\alpha$.
Alternatively, we may sample $\alpha$ conditional on the
 slab probabilities $\tau_{1}, \ldots, \tau_{H}$ through a Gibbs step (similarly to (\ref{4deq})), based on
$  \alpha|\tau_{1}, \ldots, \tau_{H} \sim \Gammad{\aalpha +H,
   \balpha - \frac{1}{H} \sum_ {h=1} ^{H}  \log \tau_{h} }$.
However, in practice we experienced that conditional
   sampling mixed poorly when compared to marginal
   sampling from $p(\alpha| H^\star )$.

\section{Application in sparse Bayesian factor analysis} \label{sec:app}

\subsection{Column-specific shrinkage of the factor loading matrix} \label{sec:SBFA}

A common application of cumulative shrinkage process priors is to identify the unknown number  of factors $\rtrue$ via the number of active columns of  the $\dimmat{\dimy}{H}$ factor loading matrix $\facload$ in an overfitting factor model with  $H>\rtrue$ potential factors, 
\begin{eqnarray}  \label{fac1reg}  
 \ym_t =  \facload  \, \facmt{t} + \errorm_t,  \qquad  \errorm_t \sim \Normult{\dimy}{\bfz,\Vare} ,
   \quad  \facmt{t}   \sim  \Normult{H}{\bfz,\identm},
\end{eqnarray}
where   $\Vare = \Diag{\sigma_1^2, \ldots, \sigma_m^2}$ is a diagonal matrix with strictly positive diagonal elements,
see \cite{roc-geo:fas,fru-etal:spa}, among many others.
To introduce
increasing column-specific shrinkage and separate the active columns of $\facload$
from the inactive ones, the  CUSP prior (\ref{CSPleg}) is applied in \cite{leg-etal:bay} in
 infinite Bayesian factor analysis where $H=\infty$ in the overfitting factor model (\ref{fac1reg}).
In the slab, a conditionally Gaussian distribution is assumed for
the elements  $\load_{ih}$ of $\facload$, with a structured prior 
variance depending on 
 a global shrinkage parameter $\tauglob$ and a column-specific shrinkage parameter $\taucol_h$,
\begin{eqnarray}
\load_{ih} | \tauglob , \taucol_h ,\idiov_i \sim \Normal{0, \tauglob  \taucol_h \idiov_i }.
   \label{priorEXP2}
   \end{eqnarray}
  The idiosyncratic variances $\idiov_i$ are assumed to follow the inverse gamma prior
   $\idiov_i \sim \Gammainv{\csigma,\bsigma}$ for all $i=1, \ldots,m$.
In recent work by \cite{sch-etal:gen}, the CUSP prior is extended to generalized infinite
 factorization models, where the factor loadings $\load_{ih}$ are allowed to be exact zeros, see also \cite{fru-etal:spa}.

 For illustration, we consider here a finite generalized
 CUSP prior on the column-specific shrinkage parameter $\taucol_h$
 in a finite overfitting model with $H \leq \maxpar{H}$,
 where $\maxpar{H} =\lfloor(m-1)/2\rfloor$ is equal to the upper bound of \cite{and-rub:sta}, ensuring econometric identification.
 We employ
 prior (\ref{priorEXP2}) for the elements $\beta_{ih}$ of the loading
 matrix and introduce a more  general spike-and-slab prior for $\taucol_h$  than the previous literature.
 More specifically,
 we assume following hierarchical ESP prior for $h=1, \ldots, H$:
\begin{eqnarray}
&  \taucol_h | S_h,\defl  \sim (1-S_h) \defl \Fd{2\acol,2\ccol} +
 S_h  \Fd{2\acol,2\ccol} , &  \label{priorNEW} \\
&  \Prob{S_h=1|\tau_h}=\tau_h, \quad \tau_h| \alpha, H \sim  \Betadis{\frac{\alpha}{H},1}.
\label{priorNEW2}  &
\end{eqnarray}
 This spike-and-slab prior for the column-specific variance parameter
 $\taucol_h$ is based on the
  F-distribution, shown in \cite{cad-etal:tri} to be a very useful prior for variance parameters. In this context it is known as the triple gamma prior.
In the spike, the shifted F-distribution
$\taucol_h|S_h=0 \sim \defl \Fd{2\acol,2\ccol}$ is assumed, where $\defl  << 1$ acts as
a deflator that pulls
the cdf of the slab toward 0. Evidently,  this prior satisfies the assumptions of
Proposition~\ref{prop41} with $\theta_0=0$:
 \begin{eqnarray*} \label{prop4Fd}
 P_{\tiny \rm slab} (\mathbb{B}_\varepsilon (0)) =
 P_{\tiny \rm slab} (\theta_h  \leq \varepsilon)=
P_{\tiny \rm spike} (\theta_h  \leq \frac{\varepsilon}{\defl} ) <
   P_{\tiny \rm spike} (\theta_h  \leq  \varepsilon) =
    P_{\tiny \rm spike} (\mathbb{B}_\varepsilon (0)).
  \end{eqnarray*}
  Prior (\ref{priorNEW}) is rather flexible and extends various shrinkage 
  priors previously suggested
in the literature.  
It has a representation as a
hierarchical mixture of inverse gamma distributions:
\begin{eqnarray} \label{priorNEW3}
&  \taucol_h | S_h, \defl, \bcol_h   \sim (1-S_h)  \Gammainv{\ccol, \defl \bcol_h} +
 S_h \Gammainv{\ccol,  \bcol_h} , \quad  \bcol_h \sim \Gammad{\acol,\acol/\ccol}. &  \end{eqnarray}
For increasing $\acol$, 
$\bcol_h$ converges to $\ccol$ and 
(\ref{priorNEW3}) is related to \cite{leg-etal:bay}. The slab distribution approaches
 $\taucol_h|S_h= 1 \sim \Gammainv{\ccol,\ccol}$,
 while the shifted spike distribution  $\taucol_h|S_h= 0 \sim \Gammainv{\ccol,\ccol \defl}$ substitutes
 the Dirac spike $\delta _{\theta_\infty}$ (where $\theta_\infty=\nu _0$) of  
  \cite{leg-etal:bay}
 with a continuous distribution with prior expectation $\defl$.
 For  $\acol=1$, prior (\ref{priorNEW})  approaches a mixture of Lasso priors
 \cite{roc-geo:fas} as $\ccol$ increases.
Finally, for $\acol=0.5$,  prior (\ref{priorNEW}) is closely related
 to the prior recently introduced
in \cite{kow-can:sem}.
For illustration, we apply  these three special cases of prior (\ref{priorNEW})  in 
\citeSec{sec:app}{illapp}.

Influential hyperparameters
of the ESP prior defined in (\ref{priorNEW})   and (\ref{priorNEW2}) such as $\alpha $, $\defl$, and $\kappa$
are learned from the data under suitable priors during MCMC sampling.
Regarding the  hyperparameter $\alpha $
 in the 1PB prior (\ref{priorNEW2}), we follow 
\citeSec{parc}{sec:hyp}
 and choose
 a gamma prior  $\alpha \sim \Gammad{\aalpha,\balpha}$ as in \cite{fru-etal:spa}.
As demonstrated by \cite{sch-can:tru}, the deflator $\defl$ in mixture (\ref{priorNEW})
can be extremely influential and has to be chosen carefully.
For this reason, we place a gamma prior with prior expectation $\Edefl << 1$ on
 this parameter, specifically $\defl \sim \Gammad{\cdefl,\cdefl/\Edefl}$. Exploiting the mixture likelihood derived from (\ref{priorNEW}) and (\ref{priorNEW2}), 
 $\defl$ is sampled from the conditional posterior
 \begin{eqnarray}  \label{Posdefl}
  p(\defl| \{\ \taucol_h, \tau_h \}_h) \propto
   p(\defl) \prod_{h=1}^H  \left[ (1-\tau_h)  p_{\tiny \rm spike} (\taucol_h|\defl)
   + \tau_h  p_{\tiny \rm slab} (\taucol_h) \right]
 \end{eqnarray}
using an MH-step.
The spike and the slab densities are easily derived from the underlying $\Fd{2\acol,2\ccol}$-distribution:
 \begin{eqnarray}  \label{classUsef}
 &&   p_{\tiny \rm slab} (\taucol_h) = 
   \frac{\acol}{\ccol\Betafun{\acol,\ccol}}
   \left(\frac{\acol \taucol_h}{\ccol}\right) ^{\acol-1}
   \left(1+\frac{\acol \taucol_h}{\ccol} \right) ^{-(\acol+\ccol)} ,\\ && \nonumber
 p_{\tiny \rm spike} (\taucol_h|\defl) =\frac{1}{ \defl} p_{\tiny \rm slab} \left(\frac{\taucol_h} { \defl}\right).
 \end{eqnarray}
The global shrinkage parameter $\kappa$   is sampled under the prior $\kappa \sim \Gammainv{\ckappa,\bkappa}$
from the conditional inverse gamma posterior
\begin{eqnarray}  \label{Poskap}
\kappa| \betav, \taucol_1, \ldots,  \taucol_H, \Vare
\sim  \Gammainv{\ckappa + \frac{m H}{2},\bkappa  + \frac{1}{2} S_{\betav}}, \quad
S_{\betav} = \sum_{h=1}^H \frac{1}{\taucol_h} \sum_{i=1}^m \frac{1}{\idiov_i} \beta_{ih}^2.
\end{eqnarray}
An important step in implementing MCMC for an ESP prior is  sampling the indicators $S_1, \ldots, S_H$ to classify the columns of the loading matrix into active and non-active ones. 
As in  \citeSec{parc}{sec:DAEPS}, classification can be based 
on the F-mixture (\ref{priorNEW})  using the  spike and slab densities
 (\ref{classUsef}):
\begin{eqnarray}  \label{sec4useF}
 \Prob{S_h =0| q_A, \theta_h,  \defl} \propto  (1- q_A) \cdot  p_{\tiny \rm spike} (\taucol_h|\defl) , \quad
 \Prob{S_h =1| q_A, \theta_h} \propto  q_A \cdot p_{\tiny \rm slab} (\theta_h) .
 \end{eqnarray}
Full  details of the MCMC procedure are given in Algorithm~\ref{Algo1}. An alternative scheme MCMC based on sampling $S_h$ conditional on $\betav_h$ and  $\bcol_h$, but marginalized w.r.t. $\theta_h$  is discussed below.

\begin{alg}[\textbf{F-classification}] \label{Algo1}
  One cycle of MCMC estimation involves the following sampling steps:\\[-0.7cm]
  \begin{itemize} \itemsep 0mm
 \item [(1)]   Sample   the model parameters
 $(\facload,\idiov_1,\ldots,\idiov_{\dimy})$ from
 $p(\facload , \idiov_1,\ldots,\idiov_{\dimy}| \taucol_1, \ldots, \taucol_H, \kappa,
   \facmt{1},\ldots,\facmt{n},\ym )$.

    For $t=1, \ldots,n$, sample the latent factors $\facmt{t}$
 from $ p(\facmt{t}|\facload,\idiov_1,\ldots, \idiov_{\dimy},\ym) $.

\item[(2)]  
Sample $\defl$ from the posterior
$p(\defl| \{\theta_h, \tau_h \}_{h=1}^H)$
  given in (\ref{Posdefl}) using a standard random walk MH step for
 $\log \defl$.

For $h=1, \ldots, H$, sample $S_h$ from the discrete posterior
(\ref{sec4useF}) using $q_A=\alpha/(\alpha+H)$.

Sample $\alpha$ from $p(\alpha| H^\star )$ given in (\ref{AMAM}) using a standard random walk MH step for
 $\log \alpha$.

For $h=1, \ldots, H$, sample $\tau_h|S_h,\alpha$ from the beta distribution (\ref{sametau}), where
$a_0=\alpha/H$ and $b_0=1$.

Sample $\bcol_h| \theta_h,\defl , S_h$  and $\theta_h| \betav_h, \Vare,\bcol_h ,\defl , S_h$ from (\ref{postADD}).

Sample $\kappa| \theta_1, \ldots, \theta_H,\betav $ from the inverse gamma distribution (\ref{Poskap}).

\end{itemize}
\end{alg}

\noindent Part~(1) of Algorithm~\ref{Algo1} encompasses standard steps in Bayesian factor analysis, see
e.g.~\cite{fru-etal:spa}. Part~(2) involves
all steps needed to implement the ESP prior introduced in  (\ref{priorNEW}) and (\ref{priorNEW2}).
The conditional posteriors
$\theta_h|\betav_h, \Vare,\bcol_h ,\defl , S_h$ and $\bcol_h| \theta_h,\defl , S_h$ are derived from
(\ref{priorNEW3}):
\begin{eqnarray} \label{postADD}
& \displaystyle \theta_h| \betav_h,\Vare, \bcol_h ,\defl , S_h \sim \Gammainv{\ccol + \frac{m}{2}, \defl^{1-S_h} \, \bcol_h +
 \frac{1}{2\kappa} \sum_{i=1}^m \frac{\beta_{ih}^2}{\idiov_i}} , &  \\
& \displaystyle \bcol_h| \theta_h,\defl , S_h \sim
 \Gammad{\acol+ \ccol,\frac{\acol}{\ccol} + \frac{\defl^{1-S_h}}{\theta_h}}. &
\nonumber
\end{eqnarray}
To enhance mixing, each cycle is concluded by a boosting step involving $\theta_1, \ldots,
 \theta_H $ and
 $\kappa$ as in  \cite{fru-etal:spa}.
An alternative MCMC scheme can be implemented which marginalizes over  $\taucol_h$ when sampling the indicators. $S_1, \ldots,S_H$.
Representation (\ref{priorNEW3})  allows to derive the conditional  spike and slab distribution 
   of the $h$th column $\betav_h$ of $\betav$ given $\bcol_h$ as  following 
    $t$-distributions: 
\begin{eqnarray} \label{classprob}
p_{\tiny \rm spike} (\betav_h|\bcol_h,\defl,\kappa,\Vare) \sim \Student{2\ccol}{\bfz,
\frac{\defl \kappa\bcol_h}{\ccol} \Vare}, \qquad
p_{\tiny \rm slab} (\betav_h|\bcol_h, \kappa,\Vare) \sim \Student{2\ccol}{\bfz,
\frac{\kappa\bcol_h}{\ccol} \Vare}.
\end{eqnarray}
These densities can be used for classification  marginalized w.r.t. $\taucol_h$:
  \begin{eqnarray}  \label{sec4at}
 &\Prob{S_h =0| q_A, \betav_h, \defl,\bcol_h, \kappa,\Vare} \propto  (1- q_A) \cdot 
p_{\tiny \rm spike} (\betav_h|\bcol_h,\defl,\kappa,\Vare), &\\\nonumber &\Prob{S_h =1| q_A, \betav_h, \bcol_h, \kappa,\Vare } \propto  q_A \cdot 
  p_{\tiny \rm slab} (\betav_h|\bcol_h, \kappa,\Vare), &
\end{eqnarray}
and to sample $\defl$ from the conditional posterior
 \begin{eqnarray}  \label{Defluset}
  p(\defl| \{\betav_h, \bcol_h, \tau_h \}_h,
\kappa, \Vare) \propto
   p(\defl) \prod_{h=1}^H  \left[ (1-\tau_h)  p_{\tiny \rm spike} (\betav_h|\bcol_h,\defl,\cdot)
+ \tau_h  p_{\tiny \rm slab} (\betav_h|\bcol_h,\cdot)
\right]
 \end{eqnarray}
using an MH-step. Furthermore, due to marginalising over $\theta_h$ rather than $\bcol_h$, the sampling order of 
 $\theta_h|\bcol_h,\cdot$ and $\bcol_h|\theta_h,\cdot$ is reversed compared to
 Algorithm~\ref{Algo1}. Fulls details of this MCMC procedure are given in Algorithm~\ref{Algo2}.

\begin{alg}[\textbf{t-classification}] \label{Algo2}
  One cycle of MCMC estimation involves the following sampling steps:\\[-0.7cm]
  \begin{itemize} \itemsep 0mm
 \item [(1)]  Same as in Algorithm~\ref{Algo1}.

\item[(2)] 
Sample $\defl$ from the posterior
given in (\ref{Defluset}) using a standard random walk MH step for
 $\log \defl$.

For $h=1, \ldots, H$, sample $S_h$ from the discrete posterior
(\ref{sec4at}) using $q_A=\alpha/(\alpha+H)$.

Sample $\alpha| H^\star$  and $\tau_h|S_h,\alpha$,  $h=1, \ldots, H$, as in Algorithm~\ref{Algo1}.

Sample $\theta_h| \betav_h, \Vare, \bcol_h ,\defl , S_h$ and $\bcol_h| \theta_h,\defl , S_h$ from (\ref{postADD}).

Sample $\kappa| \theta_1, \ldots, \theta_H,\betav $ as in Algorithm~\ref{Algo1}.

\end{itemize}
\end{alg}

\noindent 
In general, we found that Algorithm~\ref{Algo1}
exhibits better mixing than Algorithm~\ref{Algo2}, which tends to become stuck at the true value of $H_0$,  exaggerating posterior concentration; see
\citeSec{sec:app}{illapp} for illustration. 

As discussed in Section~\ref{sec:EPS}, the ESP prior (\ref{priorNEW}) and (\ref{priorNEW2}) imposes increasing shrinkage without
forcing an implicit ordering of the columns. For this reason,
Algorithm~\ref{Algo1} and \ref{Algo2} do
not impose any ordering on the columns of the loading matrix. Rather,
the ordering of the columns  remains arbitrary  during MCMC which simplifies sampling considerably.
The number of active columns $H^\star$ can be retrieved nevertheless  from the posterior draws of
$S_1, \ldots,S_H$ during MCMC, since  the functional
(\ref{CSPDA2}) is invariant to the ordering of the columns (as are several other functionals of
the posterior draws).
Increasing shrinkage becomes apparent during post-processing of the MCMC draws.
 First, we determine the  decreasing order statistics $\tau_{(1)} > \ldots > \tau_{(H)}$
 for each draw of the unordered slab probabilities
   $\tau_1, \ldots, \tau_H$ (sampled in Part~(2) of Algorithm~\ref{Algo1} or Algorithm~\ref{Algo2})
  and use the corresponding
   permutation $\rho(1), \ldots, \rho(H)$ to reorder the columns of the loading matrix.
   In this way, the generalized CUSP representation of the ESP prior is obtained, where  the sequence
   $\piL_h= 1 - \tau_{(h)}$ contains the increasing spike probabilities,
   with the corresponding column specific parameters given by
    $(\theta^\star_{1}, \ldots, \theta^\star_{H})=(\theta_{\rho(1)}, \ldots, \theta_{\rho(H)})$.
    As the column index $h$ increases,
    the marginal posterior distributions of  $\theta^\star_{h}$ and $\piL_h$
    are increasingly pulled toward   zero and  one, respectively, see Figure~\ref{fig2}
    for illustration.

\subsection{An illustrative simulation study} \label{illapp}

  \begin{figure}[t!]
\begin{center}
\scalebox{0.4}{\includegraphics{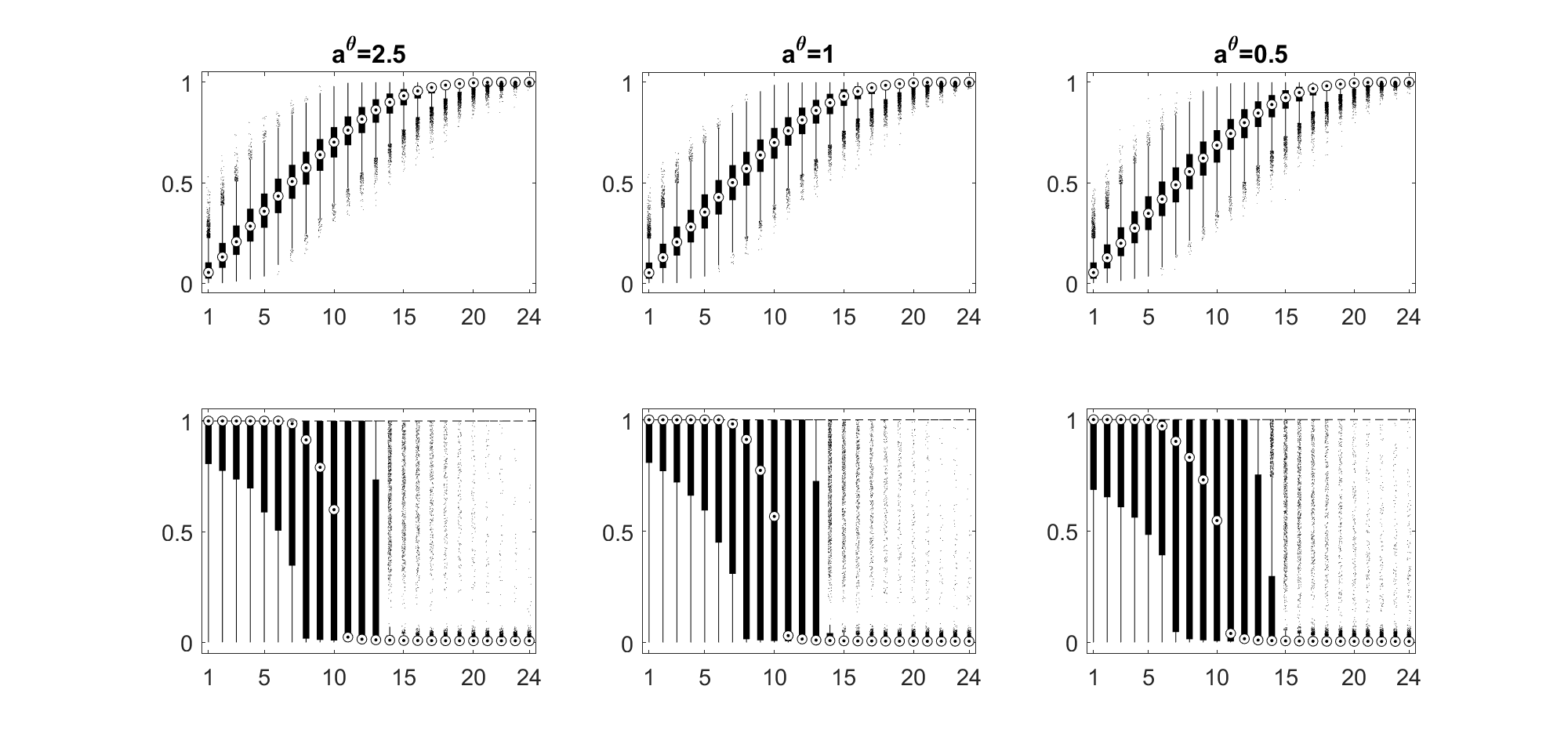}}
\caption{Data set simulated under the  dense scenario with $(m,\rtrue)=(50,10)$.
CUSP representation of the three ESP priors (from left to right),
showing  box plots of the posterior draws
of the increasing spike probability $\piL_h$ (top) and the corresponding
column specific shrinkage parameter $\theta^\star_{h}$ (bottom) for increasing
column index $h=1, \ldots,24$.}\label{fig2}
\end{center}\end{figure}

 We perform  a similar  simulation study  as \cite{leg-etal:bay}
 and consider three different combinations of $(m,\rtrue)$, namely $(20,5)$, $(50,10)$
and $(100,15)$. 
25 data sets of $n=100$ observations are sampled for 
each combination of $(m,\rtrue)$
 from the Gaussian factor model
(\ref{fac1}). In addition to the
 {\it dense} setting of \cite{leg-etal:bay}, where all  elements $\load_{ih}$ of the loading matrix $\betav_0$
 are unconstrained and drawn
 independently from
$\Normal{0,1}$, we also consider a {\it sparse} setting,
where 30\% of all $\load_{ih}=0$, while the other 70\% are drawn from a  standard normal distribution. In all
six scenarios,  $\Vare_0=\identm$.

The maximum number of active columns in the overfitting
factor model, $H=\max(\maxpar{H},30)$, increases with
$m$ and is limited to $H=30$ for $m=100$ for computational reasons.
Concerning the ESP prior on $\theta_h$, we investigate three different spike-and-slab priors in (\ref{priorNEW}): 
 an F-mixture with $\acol=2.5$,  a regularized Lasso mixture
 ($\acol=1$) and  a regularized horseshoe
 mixture ($\acol=0.5$), while $\ccol=2.5$ as 
 in \cite{leg-etal:bay} and \cite{kow-can:sem}.
Regarding the prior   on $\tau_h$ in (\ref{priorNEW2}), we learn $\alpha$ from the data under the gamma prior $\alpha \sim \Gammad{6,2}$.  This choice implies a large prior probability  
 $\Prob{\alpha << H}$ and introduces prior shrinkage on the number of active columns $H^\star$. 
 Further hyperparameter choices are
 $\csigma=2.5,\bsigma=1.5$,
  $\ckappa=\bkappa=5$,
 $\cdefl=10$,  and $\Edefl=0.01$ 
which yields, respectively,
 the prior expectations $\Ew{\sigma^2_i}=1$, $\Ew{\kappa^{-1}}=1$, and $\Ew{\defl}=0.01$.
Algorithm~\ref{Algo1} is run 
for 10,000 iterations after a burn-in of 5,000, 
starting from a
model with $H^\star=3$ active columns. Computations were implemented in MATLAB 2020 on a laptop computer with an Intel Core i5-8265U CPU with 1.60-1.80 GHz.

\begin{table}[t!] \caption{Performance of the ESP prior (\ref{priorNEW})
for 6 different data scenarios
under  $\acol=2.5$ (F),
 $\acol=1$ (L), $\acol=0.5$ (H), $\ccol=2.5$, and $\alpha \sim \Gammad{6,2}$. The columns  {\it M}  and  {\it Q} show  the median, the  5\% as well as the 95\% quantile
 of the various statistics over the simulated data sets. The results for the CUSP prior (C) are based on Table~1 of \cite{leg-etal:bay} with column {\it Q} showing the interquartile range. The last column shows the median $\tilde{s}$ of the runtime (in CPU seconds).}\label{tab1}
 \begin{center}
 \begin{tabular}{llcccccccc}  \hline
 & &  & \multicolumn{2}{c}{$\hat{H}^\star$}
   &  \multicolumn{2}{c}{$p({H}^\star=\rtrue|\ym)$}
   & \multicolumn{2}{c}{$\mbox{\rm MSE}_\Omega$}
   & \\ \multicolumn{2}{c}{$(m,\rtrue)$} &
              Prior
              &  M & Q  &  M & Q    &   M & Q  &  $\tilde{s}$\\
              \hline
     (20,5) & dense & F     & 5 & (5,5)  
     & 0.96 &  (0.87,0.98) &  0.78 &  (0.54,1.09)
& 50.2\\ &       & L     & 5 & (5,5)  & 
          0.85 &  (0.64,0.90) &   0.79 & (0.56,1.46) & 50.4  \\ 
&       & H     & 5 & (5,5)   &
          0.66 &  (0.47,0.71) &   0.87 &  (0.50,1.21)
& 51.1\\
\cline{3-10} && C &  5 & 0 & - & - & 0.75& 0.29 & 310.8 \\               \cline{2-10}
          & sparse &  F   & 5 & (5,5)  &  
          0.92 &  (0.56,0.97) &    0.50 &  (0.30,0.75)
& 51.1 \\ 
&        & L    & 5 & (5,5)  &  
          0.80 &  (0.62,0.87) &  0.46 &  (0.23,0.60) 
& 51.2 \\ 
&        &  H   & 5 & (5,5)  &
          0.55 &  (0.40,0.68) &  0.51 &  (0.27,1.16)
&  51.9\\
\hline
   (50,10) & dense & F   & 10 & (10,10)  & 
    0.98 &  (0.94,0.99) &  2.23 &   (1.60,2.95)
& 268.3  \\ &        &L    & 10 & (10,10)  &  
          0.86 &  (0.81,0.89) &  2.17 &   (1.70,3.16) 
&  271.8  \\
&        & H   & 10  & (10,10) &  
          0.57 &  (0.51,0.61) &  2.32 &  (1.75,3.26) 
& 273.6\\
\cline{3-10} && C &  10  & 0 & - & - & 2.25 & 0.33& 716.2\\         \cline{2-10}
          & sparse & F  & 10 & (10,10)   & 
          0.97 &  (0.95,0.98) &  1.16 &  (0.80, 1.61) 
& 261.5   \\ 
&        &L &    10 & (10,10)   & 
          0.81 &  (0.72,0.87) &   1.20 &  (0.87,1.66)
& 266.0 \\
&        &  H   &   10 &  (10,10) &  
     0.52 &  (0.44,0.57) &  1.17 &   (0.85,1.97)
& 266.8 \\ 
\hline
   (100,15) & dense & F & 15 & (15,15)   &  
   0.99 &  (0.98,0.99) &  3.59 &  (3.08,4.50)
& 1219.0\\ &         &L  & 15 & (15,15) & 
          0.88 &  (0.86,0.90) &  3.91 &  (3.35,4.35) 
& 1219.8 \\
&        &  H     &          15 &  (15,15) & 
     0.57 &  (0.53,0.61) &  3.81 &  (3.26,5.02)
& 1252.7\\ 
\cline{3-10} && C &  15 &  0 & - & - & 3.76 
 & 0.4 & 2284.9\\           \cline{2-10}
           & sparse &  F  & 15 & (15,15)   &  
           0.97 &  (0.96,0.99) &  1.93 &  (1.57,2.29)
& 1188.1 \\
&        &L    & 15 & (15,15) &  
          0.84 &  (0.80,0.89) &  1.97 &  (1.63, 2.34)
&  1193.8 \\ 
&        &H    & 15 &  (15,15) & 
        0.52 &  (0.40,0.56) &  2.10 &  (1.63,2.35)
& 1216.7\\ 
\hline
 \end{tabular}
 \end{center}
\end{table}

\begin{figure}[t!]
\begin{center}
\scalebox{0.4}{\includegraphics{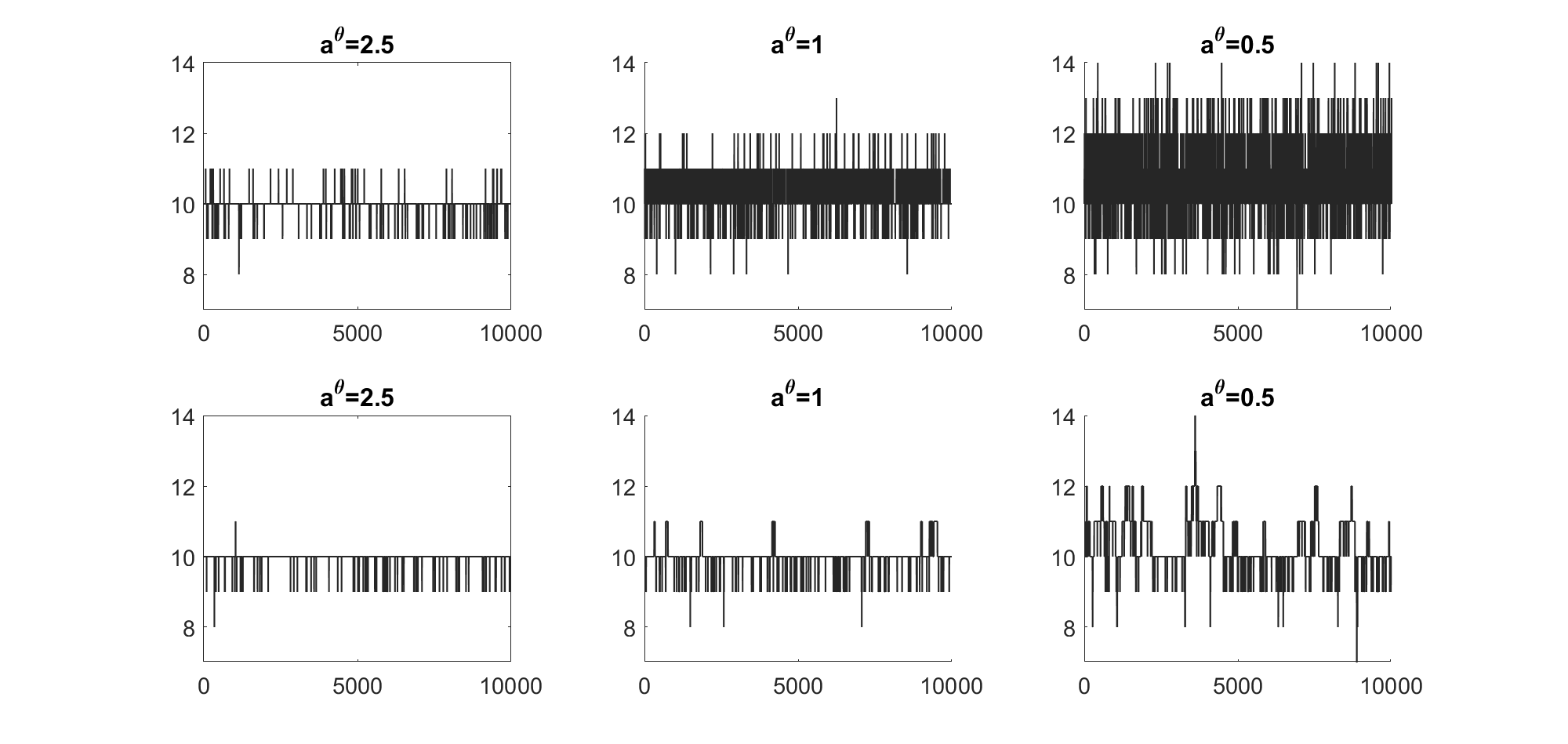}}
\caption{Data set simulated under the  dense scenario with $(m,\rtrue)=(50,10)$.
10,000 posterior draws of $H^\star$ 
using Algorithm~\ref{Algo1} (top) and
Algorithm~\ref{Algo2} (bottom) for ESP priors with  $\acol=2.5$ (left-hand side),
 $\acol=1$ (middle), $\acol=0.5$ (right-hand side), $\ccol=2.5$ and $\alpha \sim \Gammad{6,2}$.}\label{fig1}
\end{center}\end{figure}

For each of the 25 simulated data sets, we evaluate all 18 possible data
scenarios and ESP priors
through Monte Carlo estimates of following statistics:
to assess the reliability of ${H}^\star$  as an estimator of the true number $\rtrue$ of factors,
we consider the mode $\hat{H}^\star$ of the posterior distribution $p({H}^\star|\ym) $
and the magnitude of the posterior ordinate $p({H}^\star=\rtrue|\ym)$. To assess the accuracy in
estimating the true covariance matrix  $\Vary _0=\betav   _0 \betav  _0^ \top + \Vare  _0$
of the data through the  covariance matrix 
 $\Vary= \betav \betav^\top + \Vare$ implied by the overfitting model (\ref{fac1reg}), we consider the mean squared error
(MSE) defined by
\begin{eqnarray*}
\mbox{\rm MSE}_\Omega=\sum_i\sum_{\ell \leq i}
 \Ew{\left(\Omega_{i\ell}- \Omega_{0,i\ell}  \right)^2|\ym}/(m(m+1)/2).
\end{eqnarray*}
Table~\ref{tab1} reports, for all 18 possible data
scenarios and ESP priors 
the median, the 5\% as well as the 95\% quantiles of these statistics across all
25 simulated data sets. 
Regarding the covariance matrix $\Vary_0$,  all three ESP priors exhibit 
more or less  identical MSEs which  increase with $H_0$ and are  considerably smaller
 for sparse than for dense loading matrices. All three ESP priors
 are equally successful
   in recovering  $\rtrue$ from the posterior
 mode $\hat{H}^\star$. 
 Interestingly, we observe considerable variation in posterior concentration at the true value $H_0$ across  
the three values of $\acol$.
The posterior ordinate $p({H}^\star=\rtrue|\ym)$ is close to 1
  for the F-mixture with $\acol=2.5$ 
in all six data scenarios,
  even for the sparse settings. For the other two choices of $\acol$,
   $p({H}^\star=\rtrue|\ym)$ takes values considerably smaller than 1, in particular for 
    the regularized horseshoe mixture.

As explained at the end of  \citeSec{sec:app}{sec:SBFA},
the decreasing order statistics $\tau_{(1)} > \ldots > \tau_{(H)}$ of the unordered slab probabilities $\tau_1,  \ldots, \tau_H$ can be
exploited to obtain the CUSP representation of the various ESP priors.
To gain additional insights, detailed results are reported
for a  data set  simulated under the  dense scenario with $(m,\rtrue)=(50,10)$.
Figure~\ref{fig2} shows box plots of the posterior draws
of the increasing spike probability $\piL_h=1-\tau_{(h)}$  as well as the corresponding
column specific shrinkage
parameter $\theta^\star_{h}$ in the CUSP representation for all three ESP priors.
As a result of the implicit CUSP property of an ESP prior,
the posterior of the spike probability $\piL_h$
 is increasingly pulled towards one, while the
posterior of $\theta^\star_{h}$ is pulled towards zero  as $h$ increases.
Under all three ESP priors, the information in the data 
induces a clear posterior gap between active and inactive columns at the true value $h=\rtrue=10$.

Regarding MCMC performance, Algorithm~\ref{Algo1} shows good mixing
properties for all identified parameters. 
Without any thinning of the 10,000 posterior draws, the median effective sampling rate 
of, respectively,   $\log \Det{\Vary}$ and $ \|\Vary^{-1} \|_{F}$ 
across all 25 simulated data sets is, on average, equal to 27.7\% and 14.4\%, yielding an average median effective sampling size (ESS) of 2768 and 1435. 
For further illustration, Figure~\ref{fig1}
shows 10,000 posterior draws of $H^\star$
obtained by Algorithm~\ref{Algo1} and Algorithm~\ref{Algo2} for all three ESP priors 
 for a single data set simulated under the  dense scenario with $(m,\rtrue)=(50,10)$. 
Obviously, Algorithm~\ref{Algo1}, which uses the F-mixture of 
$\theta_h$ for separating active from inactive columns,
shows much better mixing than  Algorithm~\ref{Algo2}, which exploits the marginalized $t$-mixture of the columns $\betav_h$ of the loading matrix for this purpose.  
 
The variation of the posterior draws  of $H^\star$ in Figure~\ref{fig1}
  mirrors the concentration in the
posterior distribution $p(H^\star|\ym)$. 
As discussed earlier,
 the posterior
ordinate $p(H^\star=H_0=10|\ym)$ is considerably smaller than 1
 for the regularized horseshoe mixture
(see again Table~\ref{tab1}).
Under Algorithm~\ref{Algo1},  the corresponding posterior draws
show excellent mixing over the discrete posterior $p(H^\star|\ym)$, which is the main motivation
for \cite{kow-can:sem}  to suggest this prior in the first place. However, this comes at the cost of less posterior concentration.
For the regularized
Lasso mixture, posterior concentration is  more pronounced (see again Table~\ref{tab1}). Nevertheless,
the posterior draws
show rapid movement across the posterior distribution $p(H^\star|\ym)$ under Algorithm~\ref{Algo1}.
Algorithm~\ref{Algo1} yields a highly concentrated posterior distribution $p(H^\star|\ym)$ under $\acol=2.5$ 
not only for this specific data set, but also for most others (see again Table~\ref{tab1}). 

 A valid question raised by \cite{kow-can:sem} is whether such  strong posterior concentration 
is the result of a badly mixing sampler. For the specific example in Figure~\ref{fig1}, nearly perfect posterior concentration under $\acol=2.5$  is confirmed  by Algorithm~\ref{Algo2}. However, among the  450 simulated data sets we found many cases, in particular for $\acol=1$ and $\acol=0.5$, where Algorithm~\ref{Algo2} quickly moved from 
the initial model with three active columns to the true value of $H_0$, only then to get stuck at $H_0$ and produce overly optimistic posterior concentration compared to Algorithm~\ref{Algo1} (which was mixing well also in these cases).

\begin{figure}[t!]
\begin{center}
\scalebox{0.3}{\includegraphics{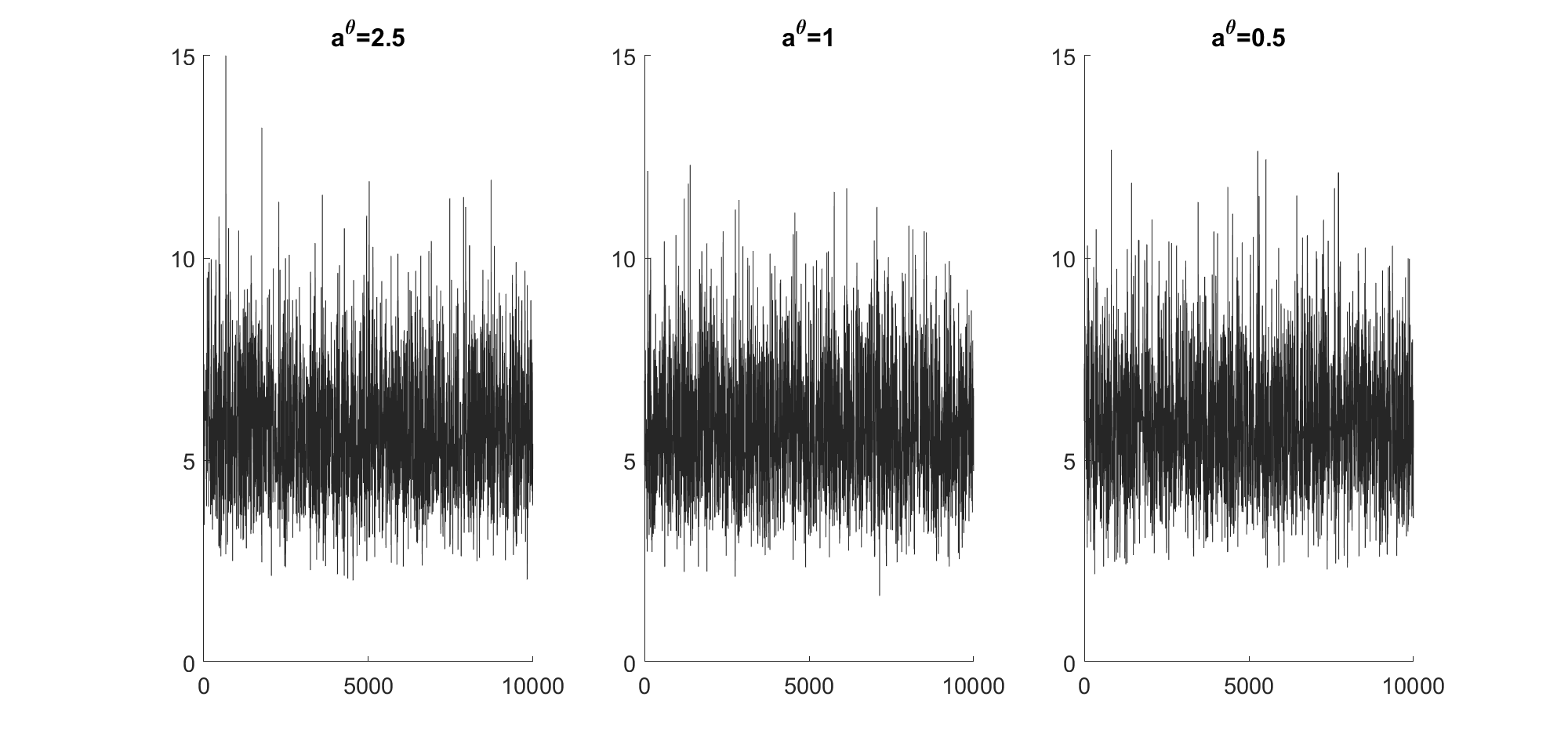}}
\caption{Data set simulated under the  dense scenario with $(m,\rtrue)=(50,10)$.
10,000 posterior draws of 
 $\alpha$  for ESP priors
with  $\acol=2.5$ (left-hand side),
 $\acol=1$ (middle), $\acol=0.5$ (right-hand side), $\ccol=2.5$ and $\alpha \sim \Gammad{6,2}$.}\label{fig3}
\end{center}\end{figure}

Finally, Figure~\ref{fig3}  
shows posterior draws of  the strength parameter of the 1BP prior, $\alpha$, 
for all three choices of $\acol$ in the spike-and-slab prior  on $\theta_h$ for the  dense scenario with $(m,\rtrue)=(50,10)$.
Posterior inference w.r.t.~$\alpha$ is rather robust regarding the choice of $\acol$ and strongly supports values of $\alpha$ considerably smaller than $H=24$. 

\paragraph*{Comparison to other priors.}
Choosing $\alpha=H$ corresponds to the uniform prior $\tau_h \sim \Uniform{0,1}$ applied in \cite{zha-etal:bay_gro} and a simplified  version of Algorithm~\ref{Algo1} with $\alpha=H$ fixed can be used for posterior inference under this prior. 
Table~\ref{tab2} shows, for all 9 combinations of sparse data scenarios and choices of 
the hyperparameter $\acol$ in the spike-and-slab prior  on $\theta_h$, how the various 
statistics change  under this prior.
 Interestingly, whether the data are informative enough to overrule the strong impact of the uniform prior on the 
 prior distribution of the 
 spike probabilities (which are pulled away from one, see again 
\citeSec{parc}{sec:hyp}) 
depends on the chosen specification for the spike-and-slab prior.  Under the F-mixture prior with $\acol=2.5$
and under the regularized Lasso mixture prior, we  still manage to retrieve the true number of factors, 
however with less posterior concentration than before. 
On the other hand, the true number of factors is systematically overfitted under the regularized horseshoe mixture prior, in particular  for the two larger models.

\begin{table}[t!] \caption{Performance of the spike-and-slab prior (\ref{priorNEW}) with 
 $\acol=2.5$ (F), $\acol=1$ (L), $\acol=0.5$ (H), and $\ccol=2.5$
 under the uniform prior $\tau_h \sim \Uniform{0,1}$  for the three sparse data scenarios.  The columns  {\it M}  and  {\it Q} show  the median, 
 the  5\% and the 95\% quantile
 of the various statistics over the simulated data sets.}\label{tab2}
 \begin{center}
 \begin{tabular}{llccccccc}  \hline
     &  & \multicolumn{2}{c}{$\hat{H}^\star$}
   &  \multicolumn{2}{c}{$p({H}^\star=\rtrue|\ym)$}
   & \multicolumn{2}{c}{$\mbox{\rm MSE}_\Omega$}  \\
 \multicolumn{1}{c}{$(m,\rtrue)$} &   
              Prior
              &   M & Q  &  M & Q    &   M & Q \\
   \hline
    (20,5) &   F   & 5 & (5,5)  & 
   0.91 &  (0.57,0.97) &  0.48 &  (0.31,0.96)   \\
                & L    & 5 & (5,5)  & 
   0.68 &  (0.28,0.78) &  0.42 &  (0.28,0.68)  \\
    &  H   & 5 & (5,6)  & 
    0.41 &   (0.32,0.51) &  0.52 &  (0.21,0.89)\\
    \hline
   (50,10) & F  & 10 & (10,10)   & 
   0.96 &   (0.89,0.98) &  1.14 &  (0.90,1.58) \\
    &L &    10 & (10,10)    & 
    0.53 &   (0.45,0.61) &  1.21 &  (0.89,1.65) \\
&H    & 12 &  (11,12) &   
0.13 &  (0.08,0.16) &  1.27 &  (0.99,1.87)  \\
 \hline
   (100,15) &   F  & 15 & (15,15)   &  
   0.98 &  (0.98,0.99) &  1.97 &  (1.69,2.61)  \\
&L    & 15 & (15,15) &  
0.62 &  (0.56,0.67) &  2.10 &  (1.64,2.47)   \\
    &H    & 17 &  (16,17) &   
    0.13  &  (0.10,0.16) &  2.00  &  (1.68,2.41) \\ 
   \hline
 \end{tabular}
 \end{center}
\end{table}

In addition, to compare finite ESP priors to the original CUSP prior, 
 we reproduce some of the performance 
measures reported in Table~1 of \cite{leg-etal:bay}
for dense factor models in Table~\ref{tab1}. The statistical performance of all finite ESP prior is identical with the CUSP prior regarding inference on $H^\star$ and 
 very similar regarding $\mbox{\rm MSE}_\Omega$. 
 However, run times improve considerably under a finite ESP prior, due to the gain in sampling the binary indicators $S_1, \ldots, S_H$ 
 instead of the categorical indicators $z_1, \ldots, z_H$. Furthermore, for the two  statistics of $\Vary$ described above,  we achieve higher effective sampling sizes than \cite{leg-etal:bay} where the median of the  ESS of 2,000 thinned draws
 is on average
 equal to 368 for a slightly different statistic of $\Vary$.  This increased sampling efficiency results from partial marginalization in Algorithm~\ref{Algo1},  where we sample
 the strength parameter $\alpha$ and the indicators $S_1, \ldots, S_H$ without conditioning on the slab probabilities $\tau_1, \ldots, \tau_H$,
 see again \citeSec{parc}{sec:DAEPS}.
 
\section{Conclusion}  \label{final}

In the present paper, we discuss shrinkage priors that
automatically impose increasing  shrinkage on a sequence of parameters.
Our main motivation  came from  Bayesian factor analysis,
where increasing  shrinkage is imposed on the loading matrix as the column index
increases to allow statistical inference with respect to
the unknown factor dimension.

We briefly reviewed
the CUSP prior of \cite{leg-etal:bay}, which is
a spike-and-slab prior, where the spike probability is stochastically increasing
 and constructed from the stick-breaking
representation of a DP prior. As a first contribution, this
prior is extended
to a generalized CUSP prior
involving arbitrary stick-breaking representations. This prior subsumes several priors introduced earlier
 in the literature,
involving various stick-breaking representations based on
beta distributions \cite{roc-geo:fas,hea-roy:gib,ohn-kim:pos,fru-etal:spa,kow-can:sem}.
As a second contribution, we prove that exchangeable spike-and-slab shrinkage (ESP) priors,
which are popular
and widely used in many areas of applied  Bayesian inference,
can be represented as
a finite generalized  CUSP prior. The CUSP representation can be easily
derived from the decreasing order statistics
of the slab probabilities.

Working with an ESP prior on a sequence of parameters which is invariant
 to the ordering and, at the same time,
implicitly imposes increasing shrinkage
without forcing explicit order constraints on the slab probabilities
is very convenient.
 It allows, in particular, to design efficient MCMC samplers under the ESP prior
and to derive the CUSP representation during post-processing. As opposed to this,
direct sampling under the order constraints in the CUSP representation is more challenging
and often a truncated CUSP prior with $H < \infty$ has to be
 employed for infinite models. Using instead
an ESP prior with large $H$ with the same spike-and-slab distribution for
the parameter $\theta_h$ as the infinite CUSP prior will induce similar increasing shrinkage, while
 classification is much simplified and reduces to sampling $H$ binary indicators instead of
     $H$ categorical variables with $H$ categories.

An application to  sparse Bayesian factor analysis
illustrates the usefulness of the findings of this paper.
  A new exchangeable  spike-and-slab shrinkage prior based on the triple gamma prior
\cite{cad-etal:tri} is introduced. In the context of Bayesian factor analysis,
 this ESP prior induces  increasing
shrinkage in the columns of the loading matrix.
In a simulation study it is shown that this prior is helpful
for estimating the unknown number of factors.
The main focus of this application to sparse Bayesian factor analysis
 lies on column sparsity, but as mentioned
in the introduction,  element-wise sparsity is another common goal in factor analysis.
Combining both approaches is an interesting venue for further research and is investigated
in \cite{fru-etal:spa}.

\section*{Appendix}

We prove that the finite 1PB prior  (\ref{prialt1P})
  has a representation as a finite  CUSP prior as in  (\ref{CSP2}) and (\ref{CSP1}),
  where the sticks $\stick_h$,
      $h=1, \ldots, H$
      are an independent sequence of beta random variables,
     \begin{eqnarray}  \label{CSP6ext}
\stick_h   \sim  \Betadis{1,\alpha \frac{H-h+1}{H}}.
 \end{eqnarray}
  Equivalently, the decreasing slab probabilities can be represented as:
    \begin{eqnarray} \label{CSP6}
 \tau_{(h)}= 1 - \piL_h = \prod _{\ell =1}^{h} (1-\stick_\ell)= \prod _{\ell =1}^{h} \nu^\star_\ell
 = \tau_{(h-1)} \nu^\star_h,
 \end{eqnarray}
 where $\nu^\star_h \,\, i.i.d. \, \, \Betadis{\alpha\frac{H-h+1}{H} ,1}$.
It is easy to show that $\Ew{\piL_h} $  satisfies following recursion for all $h$.
 From (\ref{CSP6}) we obtain
  \begin{eqnarray*} \Ew{\tau_{(h)}}=  \Ew{\tau_{(h-1)}} \Ew{\nu^\star_h} =  \Ew{\tau_{(h-1)}} C_h,
\qquad  C_h= \frac{\alpha(1-\frac{h-1}{H})}{\alpha(1-\frac{h-1}{H})+1} < 1.
 \end{eqnarray*}
Hence, $\Ew{\tau_{(h)}}$ is a decreasing sequence and, consequently,
$\Ew{\piL_h} = 1 - \Ew{\tau_{(h)}}$ is increasing.
It follows immediately that  prior  (\ref{prialt1P})
converges to the
 CUSP prior proposed by \cite{leg-etal:bay} with
 strength parameter $\alpha$ as $H$ goes to infinity,
since $\lim_{H\rightarrow \infty} \frac{H-h+1}{H}= 1$.
To prove (\ref{CSP6ext}) and (\ref{CSP6}), we follow \cite{teh-etal:sti} and
   take a closer look at the distribution of the decreasing order statistics
   $\tau_{(1)} > \ldots > \tau_{(H)}$.
  The unordered slab probabilities $\tau_h$ in (\ref{prialt1P}) exhibit, respectively,
  following pdf and cdf:
  \begin{eqnarray*}
 p(\tau_h)=\frac{\alpha}{H}  \tau_h^{\frac{\alpha}{H}-1}, \quad
 F_{\tau_h} (\tau)= \Prob{\tau_h \leq \tau}= \tau^{\frac{\alpha}{H}}.
 \end{eqnarray*}
 Let $T$ be an arbitrary natural number.
  First we show that for any  sequence of $T$ iid  r.v. $X_h \sim \Betadis{\frac{\alpha}{H},1}$,
  the  maximum
$ \maxpar{X} =  \max (X_1, \ldots, X_T)$ follows the  $\Betadis{\frac{T\alpha}{H},1}$-distribution:
 \begin{eqnarray} \label{norderst}
F_{X_{\max}} (x)=
  \prod_{h=1}^T \Prob{X_h \leq x}= (x^{\frac{\alpha}{H}})^T= x^{\frac{T\alpha}{H}}.
 \end{eqnarray}
With $T=H$, it follows from  (\ref{norderst}) that
under prior (\ref{prialt1P}) the largest order statistic,
$\tau_{(1)}= \max (\tau_1, \ldots, \tau_H) $  follows
$\tau_{(1)} \sim  \Betadis{\alpha,1} $.
Given the order statistic $\tau_{(1)}, \ldots, \tau_{(h-1)}$, the range the
remaining unordered slab probabilities  $\tau_h| (\tau_h<\tau_{(h-1)})$ is obviously restricted to $[0, \tau_{(h-1)}]$ and the corresponding cdf
 is given by:
\begin{eqnarray*}
  F_{\tau_{h}} (\tau) =  \Prob{\tau_h \leq \tau  |  \tau_h< \tau_{(h-1)}  }=
  \frac{\int_0 ^{\tau } \frac{\alpha}{H}  \tau_h ^{\frac{\alpha}{H}-1} \, d \, \tau _h }
  {\int_0 ^{\tau_{(h-1)}}  \frac{\alpha}{H}  \tau_h^{\frac{\alpha}{H}-1} \, d \, \tau _h }
  = \frac{\tau ^{\frac{\alpha}{H}}}{(\tau_{(h-1)})^{\frac{\alpha}{H}}} =
  \left(\frac{\tau}{\tau_{(h-1)}}\right)^{\frac{\alpha}{H}}.
   \end{eqnarray*}
 Hence, all $H-h+1$ slab probabilities
 $\tau_h$ smaller than $\tau_{(h-1)}$ are independent and can be presented as
$\tau_h = X_h \tau_{(h-1)}$, were $X_h \sim  \Betadis{\frac{\alpha}{H},1}$.
This follows immediately from
   \begin{eqnarray*}
  F_{X_h} (x) =   \Prob{X_h \leq x}=
\Prob{\tau_h \leq x \tau_{(h-1)} }=
  \left(\frac{ x \tau_{(h-1)}}{\tau_{(h-1)}}\right)^{\frac{\alpha}{H}}=x^{\frac{\alpha}{H}}.
   \end{eqnarray*}
  Given the order statistic $\tau_{(h-1)}$, the order statistic $\tau_{(h)}$ can be derived
   as $\tau_{(h)}= \tau_{(h-1)} \stick_{h}^\star$, where $\stick_{h}^\star=
  \max_{\ell: \tau_\ell < \tau_{(h-1)} } X_h$. Using (\ref{norderst}) with $T=H-h+1$, we obtain:
   \begin{eqnarray*}
 \tau_{(h)}= \tau_{(h-1)} \stick_{h}^\star , \qquad
  \stick_{h}^\star \sim  \Betadis{\alpha \frac{H-h+1}{H},1}.
 \end{eqnarray*}
 From this, it follows immediately that the ordered spike probabilities
  $\piL_h= 1 - \tau_{(h)}$ and the ordered slab probabilities $\piL_h^\star=\tau_{(h)}$
can be represented as, respectively, in (\ref{CSP6ext}) and (\ref{CSP6}).
 
\bibliographystyle{chicago}
\bibliography{sylvia_kyoto}

\end{document}